\documentclass[aps,prb,twocolumn,showpacs,floats]{revtex4}

\usepackage{graphicx}
\usepackage{amsmath}
\graphicspath{ {Images/} }
\DeclareGraphicsExtensions{.pdf,.png}

\newcommand{\be}{\begin{equation}}
\newcommand{\ee}{\end{equation}}
\newcommand{\bea}{\begin{eqnarray}}
\newcommand{\eea}{\end{eqnarray}}
\newcommand{\ba}{\begin{array}}
\newcommand{\ea}{\end{array}}
\newcommand{\nn}{\nonumber}

\newcommand{\Del}{\Delta}

\newcommand{\al}{\alpha}

\newcommand{\noi}{\noindent}

\newcommand{\ra}{\rangle}
\newcommand{\la}{\langle}

\begin{document}

\title{\bf Effect of transverse anisotropy on inelastic tunneling spectroscopy of atomic-scale magnetic chains}
\author{J. Hageman} \email{jarthageman@gmail.com}
\author{M. Blaauboer} \email{m.blaauboer@tudelft.nl}
\affiliation{Kavli Institute of Nanoscience, Delft University of Technology, Lorentzweg 1. 2628 CJ Delft, The Netherlands}

\date{\today}

\begin{abstract}
We theoretically investigate the effect of transverse magnetic anisotropy on spin-flip assisted tunneling through atomic spin chains. 
Using a phenomenological approach and first-order perturbation theory, we analytically calculate the inelastic tunneling current, 
differential conductance and atomic spin transition rates. We predict the appearance of additional steps in the differential conductance
and a pronounced increase in the spin-flip transition rate which at low voltages scale quadratically with the ratio of the transverse anisotropy energy 
and the sum of the longitudinal anisotropy energy and the exchange energy. 
Our results provide intuitive qualitative insight in the role played by transverse anisotropy in inelastic tunneling spectroscopy of atomic chains and can be observed under 
realistic experimental conditions.
\end{abstract}

\maketitle

\section{Introduction}

The development of small electronic storage devices has been going on for years and
the possibility to store information in magnetic nanostructures is an active topic of research. 
A recent experiment has shown that information can in principle be stored in just a few 
antiferromagnetically aligned Fe atoms by using exchange coupling between atomic spins~\cite{loth12}. 
This experiment involved imaging and manipulation of individual atomic spins by spin-polarized 
scanning tunneling microscopy (STM), a technique which over the last decade has developed into a 
powerful tool for studying spin dynamics of engineered atomic structures. In a series of seminal STM experiments 
inelastic tunneling spectroscopy (IETS) has been used to investigate 
spin excitation spectra of individual magnetic atoms~\cite{hein04}, to probe the exchange interaction between spins
in chains of Mn atoms and the orientation and strength of their magnetic anisotropy~\cite{hirj06,hirj07}, 
and to study the effect of this anisotropy on Kondo screening of magnetic atoms~\cite{otte08}. A few years later, experimental studies of 
tunneling-induced spin dynamics in atomically assembled magnetic structures were performed: In 2012 Loth {\it et al.} 
measured the voltage-induced switching rate between the two N\'eel ground states of an antiferromagnetically coupled chain 
of Fe atoms~\cite{loth12} and recently spin waves (magnons) have been imaged in chains of ferromagnetically aligned atoms, including the demonstration of 
switching between the two oppositely aligned ground states and
local tuning of spin state mixing by exchange coupling~\cite{spin14,khat13,yan15}.
\vspace*{0.2cm} \\
The tunnel-current-induced spin dynamics of single magnetic atoms and engineered atomic chains in these experiments can be well described
by a spin Heisenberg Hamiltonian (see section~\ref{sec-Hamiltonian}), 
which contains the magnetic anisotropy and exchange coupling between neighbouring atomic spins as phenomenological parameters. 
This model has been succesfully 
used to analyze, among others, the $I(V)$ characteristics of an electron interacting via exchange coupling 
with a magnetic atom~\cite{hirj06,hirj07,otte08,fern09}, to explain step heights in inelastic conductance 
measurements of adsorbed Fe atoms~\cite{lore09,soth10}, to provide a theoretical description based on rate equations of 
spin dynamics in one-dimensional chains of magnetic atoms~\cite{delg10}, to analyze magnetic switching in terms of 
the underlying quantum processes in ferrromagnetic chains~\cite{gauy13} and to calculate the 
electron-induced switching rate between N\'eel states in antiferromagnetic chains of Fe atoms~\cite{gauy13PRL,li15,tern15}.
\vspace*{0.2cm} \\
In this paper we investigate the effect of single-spin transverse magnetic anisotropy on spin-flip assisted tunneling and spin transition rates in chains
of magnetic atoms.
Understanding the role played by magnetic anisotropy in tunnelling spectroscopy is of great importance both fundamentally 
and for being able to engineer magnetic properties of atomic chains and clusters on surfaces, as well as those of molecular magnets~\cite{gatt06,burz15}.
Compared with the longitudinal (easy-axis) magnetic anisotropy, the qualitative effect of transverse magnetic anisotropy on tunneling spectroscopy of 
magnetic chains has so far received little attention. In experiments involving antiferromagnetically coupled atoms  
transverse anisotropy has often been small (i.e. too small to be observable) to negligible, 
because the easy-axis anisotropy energy is much larger than the transverse exchange energy~\cite{hirj06,hirj07,otte08,loth12}. 
However, such a uni-axial model does not always apply. Transverse anisotropy, together with the parity of the atomic spin, influences the degeneracy
of the energy spectrum~\cite{hirj07,delg12,jaco15}. Recent studies have demonstrated that the presence of 
non-zero transverse anisotropy modifies the switching frequency of few-atom magnets when atoms are directly adsorbed on the substrate~\cite{khat13}. 
It has also been predicted that finite values of transverse anisotropy lead to the appearance of peaks in the 
differential conductance when using spin-polarized STM~\cite{misi13} and a recent experiment has 
demonstrated that the strength of the magnetocrystalline anisotropy can be controllably enhanced or reduced by manipulating its local strain environment~\cite{brya13}. 
In addition, ferromagnetically coupled atomic chains (nanomagnets) usually 
exhibit non-negligible values of transverse anisotropy~\cite{spin14,khat13,yan15}.
From an engineering point of view, transverse anisotropy could be used to tune  dynamic properties such as spin switching 
in antiferromagnetic chains, since it breaks the degeneracy of the 
N\'eel ground states and transforms them into N\'eel-like states that contain a larger number of different spin configurations~\cite{switching_vs_E}.
Recent experiments have investigated the three-dimensional distribution of the magnetic anisotropy of single Fe atoms and demonstrated
the electronic tunability of 
the relative magnitude of longitudinal and transverse anisotropy~\cite{yan15_2}. This provides
further evidence for the potential importance of tunability of magnetic anisotropy for enhancing or weakening spin tunneling phenomena in magnetic adatoms and molecular 
magnets~\cite{ober14,burz15}. Given all this, 
it is interesting and important to obtain direct and intuitive qualitative insight 
in the effect of transverse anisotropy on inelastic tunneling transport and STM-induced spin 
transition rates in chains of magnetic atoms. The aim of this paper is to provide a first step in this direction on a phenomenological level.
\vspace*{0.2cm} \\
Using a perturbative approach and including the strength of the transverse anisotropy up to first order, we analytically calculate the inelastic current $I(V)$, 
differential conductance $dI/dV$ and corresponding IETS spectra $d^2I/dV^2$ for atomic chains with nearest-neighbour Ising exchange coupling. We also 
perform numerical simulations of spin transition rates of an
antiferromagnetically coupled atomic spin chain. We find that finite transverse anisotropy introduces: 1) additional steps in the differential
conductance $dI/dV$ and corresponding sharp peaks in $d^2I/dV^2$ and 2) a 
substantial increase of the spin transition rate between atomic levels.
We show that both are due to transverse anisotropy-induced coupling between additional atomic spin levels and provide a qualitative explanation of the 
position and heights of the conductance steps and the dependence of the spin transition rates on the strength of the transverse anisotropy. 
Our perturbative approach is valid for single-spin transverse anisotropy strengths 
corresponding to typical experimental values.
\vspace*{0.2cm} \\
The outline of the paper is as follows. In Sec.~II A we discuss the phenomenological spin Hamiltonian and its energy spectrum and eigenfunctions 
with the transverse anisotropy energy included up to first-order perturbation theory. We then derive analytical expressions for the inelastic tunneling 
current $I(V)$, differential conductance $dI/dV$ and IETS spectra $d^2I/dV^2$ of an $N$-atomic spin chain (Sec. II B),
and for the tunneling-induced transition rates (Sec. II C). Application of these results to chains of antiferromagnetically coupled atoms
are presented and analyzed in Secs.~III and IV. Sec.~V contains conclusions and a discussion of open questions.
\section{Theory}
\subsection{Hamiltonian}
\label{sec-Hamiltonian}
\begin{figure}[h]
\includegraphics[width=0.5\textwidth]{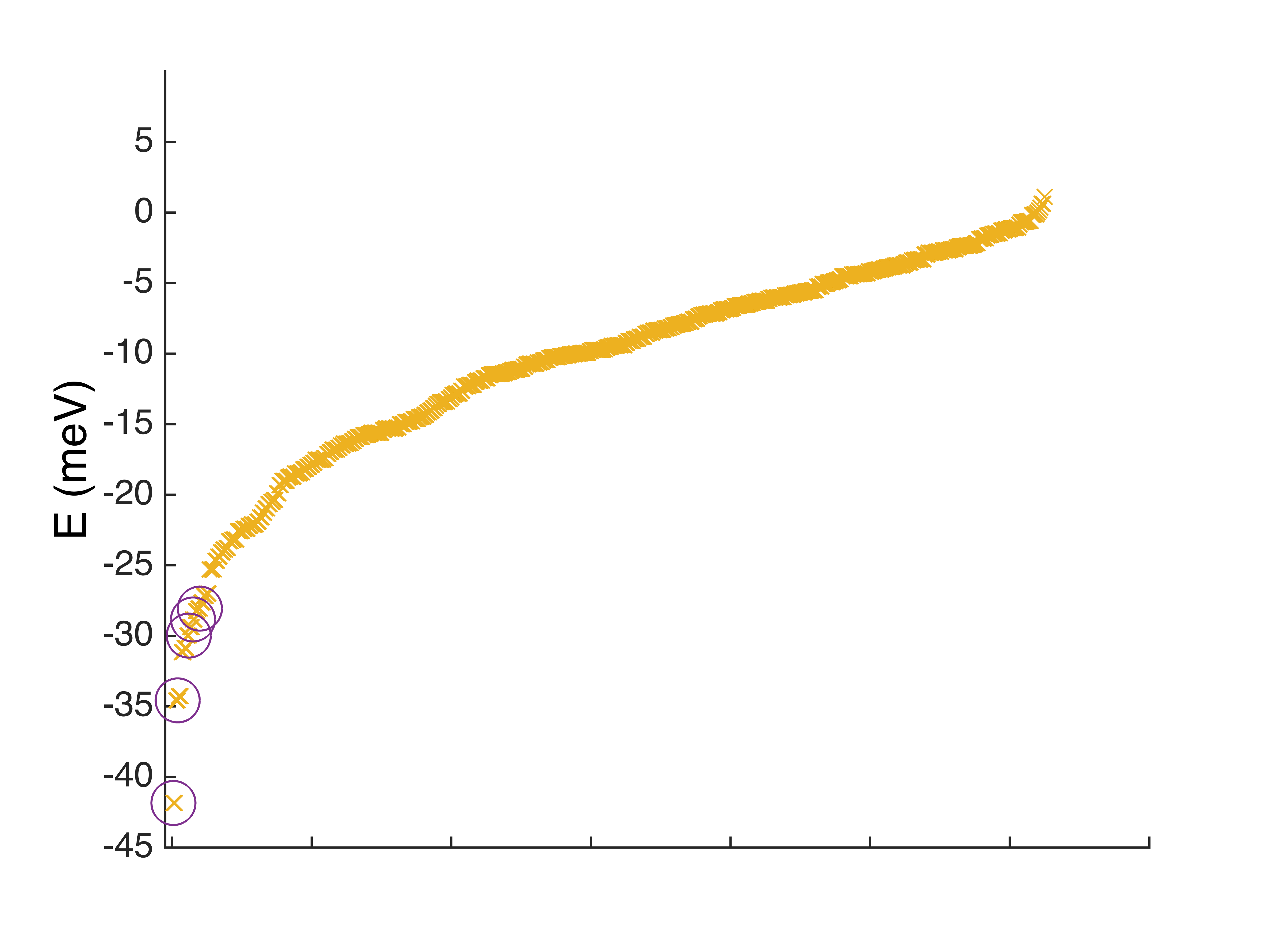}
\caption{The energy spectrum $E\equiv E_{m_1,..,m_N}$ [Eq.~(\ref{eq:energies})] of an antiferromagnetically coupled chain consisting 
of four Fe atoms ($s=2$). Each cross represents an eigenstate, plotted in ascending order of energy. The circles mark the five eigenstates that contribute to the current
and spin transition rates in Figs.~\ref{fig:IVAFM}-\ref{fig:PM}.
Parameters used are $D=-1.3$\,meV, $J=1.7$\, meV and $B=1$\,T.}
\label{fig:AFMspec2D}
\end{figure}

In this section we first briefly discuss the spin Hamiltonian used to describe the atomic chain and then derive its eigenvalues and 
eigenfunctions up to first order in the strength of the transverse magnetic anisotropy.
\\
The eigenenergies and spin eigenstates of a chain of $N$ magnetic atoms can be described by a phenomenological Heisenberg spin Hamiltonian, 
consisting of a single-spin part and nearest-neighbour exchange interaction~\cite{hirj06,spin14,fern09,delg10,gatt06}:
\be
{\cal H} =  \sum_{i=1}^{N} \hat{\cal H}_{i,S} + \sum_{i=1}^{N-1} J\, \hat{\bf S}_i \cdot \hat{\bf S}_{i+1} 
\label{TotalHamiltonian}
\ee
with
\be
\hat{\cal H}_{i,S}  = D\hat{S}^2_{i,z} + E(\hat{S}^2_{i,x} - \hat{S}^2_{i,y}) - g^*{\mu_B} {\bf B} \cdot \hat{\bf S}_{i}.
\label{SinglespinHamiltonian}
\ee
Here $D$ represents the single-spin longitudinal magnetocrystalline anisotropy, $E$ the transverse magnetic anisotropy, $g^*$ the Land\'{e} g-factor, 
$\mu_B$ the Bohr-magnetron, and ${\bf B}$ the external magnetic field. $J$  denotes the exchange energy between neighbouring atoms, which
can be directionally dependent with different energies $J_x$, $J_y$ and $J_z$ for the three spin directions.
In principle, $D$, $E$ and $J$ depend on the substrate and can vary from atom to atom. However, since experimentally the atom-to-atom variations 
are found to be small, see e.g. Refs.~\cite{loth12,spin14}, these coefficients can to a good approximation be assumed to be uniform along the chain. 
$\hat{S}_{i,x}$, $\hat{S}_{i,y}$ and $\hat{S}_{i,z}$ are, respectively, the $x$-, $y$-, and $z$-components of the spin operator of the atom at site $i$
along the chain. Assuming ${\bf B} = B \hat{z}$, $E=0$ and exchange coupling between the $z$-components of the spin only (Ising coupling, see 
also the end of this section), the eigenvalues $E_{m_1,..., m_N}$ of the Hamiltonian~(\ref{TotalHamiltonian}) are given by
\be
E_{m_1,..., m_N} = D \sum_{i=1}^{N} m_i^2 + J \sum_{i=1}^{N-1} m_i m_{i+1} - E_Z \sum_{i=1}^{N} m_i,
\label{eq:energies}
\ee
with corresponding eigenstates $|m_1,..m_N\ra^{(0)}$. Here $m_i$ denotes the quantum number labeling the angular momentum in the $z$-direction 
of the $i^{th}$ atom and $E_Z\equiv g^{*} \mu_B B$ represents the Zeeman energy. 
When adding the single-spin transverse anisotropy
$E(\hat{S}^2_{i,x} - \hat{S}^2_{i,y})$ as a perturbation, the corresponding eigenfunctions $\psi_{m_1,..,m_N}$ up to first order in $E$ are given by
\be
\psi_{m_1,\ldots,m_N} = |m_1,..,m_N\ra^{(0)} +  |m_1,..,m_N\ra^{(1)},
\label{eigenstate-gestoord}
\ee
with
\bea
|m_1,...m_N\ra^{(1)} & = & 
 \frac{1}{2} E \, \sum_{i=1}^{N} \nn \\ 
 & & \left( \right. A_{m_{i-1},m_i,m_{i+1}} |m_1..m_i\! +\!  2..m_N\ra^{(0)}\, - \nn \\
& & \left. A_{m_{i-1},m_i-2,m_{i+1}} |m_1..m_i\!  - \!  2..m_N\ra^{(0)} \right) \nn \\
\label{eq:corr_psi_1}
\eea
and
\bea
A_{m_{i-1},m_i,m_{i+1}}  & \equiv  & \nn \\
\left( 36 - 12 (m_i + 1)^2  \right. & + & \left.
 m_i (m_i + 1)^2 (m_i + 2) +36 \right)^{\frac{1}{2}}/ \nn \\
\left( - 4D(m_i + 1) \right. & - & \left. 2J (m_{i-1} + m_{i+1}) + 2 E_Z \right).
\label{eq:Ami} 
\eea
Here $A_{m_{j-1},m_j,m_{j+1}} = 0$ for $j< 1$ or $m_{j-1},m_j,m_{j+1} \notin [-2,..,2]$, $A_{m_0,m_1,m_2} \equiv A_{0,m_1,m_2}$ and 
$A_{m_{N-1},m_N,m_{N+1}} \equiv A_{m_{N-1},m_N,0}$. 
Since the first-order correction of the eigenenergies~(\ref{eq:energies}) is zero, Eqns.~(\ref{eq:energies}) and (\ref{eigenstate-gestoord}) 
thus represent the eigenvalues and eigenfunctions of the Hamiltonian (\ref{TotalHamiltonian})
(for ${\bf B} = B \hat{z}$ and Ising coupling) up to first order in $E$.
\\
Figure~\ref{fig:AFMspec2D} shows the energy spectrum (\ref{eq:energies}) for a chain consisting of four Fe atoms (spin $s=2$) with antiferromagnetic coupling. 
The ground state of the chain consists of the two degenerate N\'eel states $|2,-2,2,-2\ra^{(0)}$ and $|-2,2,-2,2\ra^{(0)}$ with eigenenergy $16 D - 12 J$. 
\\
\noi We end this section with a brief discussion of the assumption of Ising coupling in the Hamiltonian (\ref{TotalHamiltonian}). This assumption is in 
general a good approximation for the description of atomic chains in which the longitudinal anisotropy is at least of the same order of
magnitude as the exchange coupling, $|D/J| \gtrsim 1$, such as those studied in Refs.~\cite{loth12,yan15}.
When including the transverse anisotropy term in the Hamiltonian, which contains the off-diagonal spin operators $\hat{S}_x$ and $\hat{S}_y$, the question
arises whether these non-Ising magnetic exchange terms pose additional requirements on the validity of the Ising approximation. 
We expect this approximation to be valid for atomic chains with predominantly longitudinal exchange coupling 
$|J_{x,y}/J_z| << 1$ (where the eigenstates are to a good approximation the eigenstates of the Ising model),
and transverse anisotropy strengths $|J_{x,y}/E| \ll 1$ (so that $E$ is the dominant energy scale in the off-diagonal terms). 
\subsection{Current}
\label{sec-current}
A powerful technique to probe the spin dynamics of single magnetic atoms or small atomic chains deposited on a surface (typically a thin insulating layer on top of a metallic surface) is 
inelastic tunneling spectroscopy (IETS). In IETS, the spin of an electron tunneling from the tip of an STM interacts via exchange with the spin of an atom. When the energy provided 
by the bias voltage matches the energy of an atomic spin transition, the latter can occur and a new conduction channel opens~\cite{jakl66}. 
For a chain consisting of $N$ magnetic atoms with spin $s=2$ (such as Fe or Mn) and the STM tip located above atom $j$ the inelastic current $I(V)$ in an IETS experiment is 
given by
\bea
I(V) & = & G_S\, \prod_{i=1}^N \, \left( \sum_{\stackrel{m_i,m_i^{\prime}=-2}{\al=x,y,z;s=\pm}}^{2} \! \right) \, P_M(V) \times \nn \\
& & |\la m_1,..,m_N | S_{\al}^{(j)} | m_1^{\prime},..,m_N^{\prime} \ra |^2\, F_{M,M^{\prime}\!,s}(V)
\label{eq:I_V}
\eea
with 
\be
F_{M,M^{\prime},s}(V) \equiv \frac{eV - s \Del_{M^{\prime}\!,M}}{ 1 - e^{- s\beta (eV - s\Del_{M^{\prime}\!,M}
)} }.
\label{eq:FactorF}
\ee
Eq.~(\ref{eq:I_V}) is the $N$-atom generalization of the expression for the current given in Ref.~\cite{fern09}. 
Here $| m_1,..,m_N \ra$ and $| m_1^{\prime},..,m_N^{\prime} \ra$ denote the eigenstates of the Hamiltonian (\ref{TotalHamiltonian}), 
$M$ represents the set of quantum numbers $(m_1,..m_N)$ corresponding to the eigenstate $|m_1,..,m_N\ra^{(0)}$ of the Hamiltonian (\ref{TotalHamiltonian})
for ${\bf B} = B \hat{z}$, $E=0$ and Ising coupling, and $\Del_{M^{\prime}\!,M} \equiv E_{M^{\prime}} - E_{M}$ with $E_M$ given
by Eq.~(\ref{eq:energies}). $P_M(V)$ is the occupation of eigenstate $\psi_{m_1,..,m_N}$ (see also the appendix), $\beta \equiv (k_B T)^{-1}$ 
and $S_{\al}^{(j)}$ with $\al=x, y, z$ 
is the local spin operator acting on atom $j$. The conductance quantum $G_S \sim \frac{2e^2}{h} \rho_T \rho_S T_S^2$ with 
$\rho_T, \rho_S$ the density of states at the Fermi energy of the STM tip and surface electrodes 
and $T_S^2$ the tunneling probability between the local atomic spin and the transport electrons~\cite{delg10}. Eq.~(\ref{eq:I_V})
has been derived (for a single atomic spin) starting from a microscopic tunneling Hamiltonian that describes the exchange interaction 
between the spin of the tunneling 
electron and the atomic spin assuming short-range exchange interaction~\cite{fern09} (alternative approaches that have been 
used to study spin-flip assisted transport in chains of magnetic atoms include nonequilibrium Green's functions~\cite{fran10} 
and generalized Anderson models~\cite{delg11}). The STM tip then only couples to the atom at site $j$ and the matrix element 
$|\la m_1,...,m_N | S_{\al}^{(j)} | m_1^{\prime},...m_N^{\prime} \ra |^2$ describes the exchange spin interaction between the spin 
of the tunneling electron and this atomic spin: the generalization for coupling to several atoms is 
$|\la m_1,..,m_N | \vec{S}_{\al} | m_1^{\prime},..,m_N^{\prime} \ra |^2$ with $\vec{S}_{\al} = \sum_j \eta(j) S_{\al}^{(j)}$ and 
$\eta(j)$ the tunnel probability through atom $j$, see Ref.~\cite{fern09}. The function $F_{M^{\prime}\!,M,s}(V)$ on the right-hand 
side of Eq.~(\ref{eq:I_V}) is the temperature-dependent activation energy 
for opening a new conduction channel: at energies where the 
applied bias voltage $eV$ matches the energy that is required for an atomic spin transition $\Del_{M^{\prime}\!,M}$ a step-like increase in the differential conductance $dI/dV$ occurs. 
At these same voltages the second derivative of the current $d^2I/dV^2$ exhibits a peak 
(of approximately Gaussian shape). The area under these peaks corresponds to the relative transition intensity and is equal to the corresponding step height of the differential conductance~\cite{yan15_2}. 
Analyzing $d^2 I/dV^2$ data, commonly called IETS spectra, thus probes the transition probability between atomic spin levels.~\cite{spin14,yan15}

\begin{figure}[h]
\includegraphics[width=0.5\textwidth]{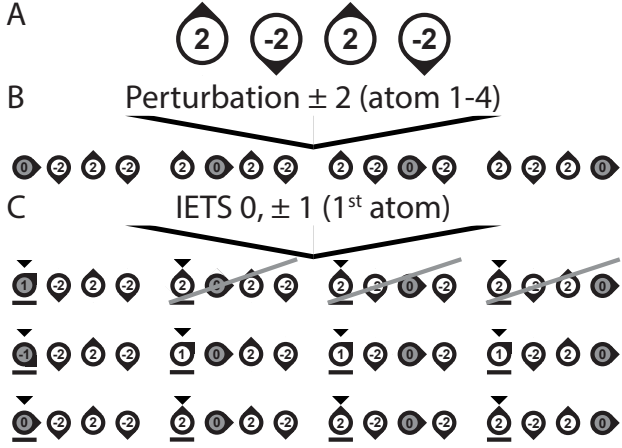}
\caption{Schematic illustration of the eigenstates contributing to the IETS current (\ref{eq:current}) for a four-atom chain
in the ground state. Panel A: The unperturbed ground state $|2,-2,2,-2\ra^{(0)}$. Panel B: The four unperturbed eigenstates 
contributing to the new ground state after including the transverse anisotropy energy to first order
in perturbation theory. In each of these eigenstates $m_i$ differs by $-2$ (for $i=1,3$) or $+2$ (for $i=2,4$) from the unperturbed ground state in panel A (see also Eq.~(\ref{eigenstate-gestoord})). 
Panel C: The unperturbed eigenstates contributing to the IETS current if the STM tip is coupled
to the first atom. In these eigenstates $m_1$ differs by $+1$ (upper row), $-1$ (middle row) or $0$ (lower row) compared to the states in panel B.}
\label{fig:scheme}
\end{figure}

We now calculate the inelastic current (\ref{eq:I_V}) for a nonmagnetic STM tip by calculating the spin exchange matrix element for the eigenstates 
Eq.~(\ref{eigenstate-gestoord}), i.e. using $|m_1,..,m_N\ra = \psi_{m1,..,m_N}$,
and the occupation probabilities $P_M(V)$ for each eigenstate $|m_1,..,m_N\ra^{(0)}$. 
These populations $P_M(V)$ are obtained by solving the master equation~\cite{delg10}
\bea
\frac{dP_M(V)}{dt} & = & \sum_{M^{\prime}} P_{M^{\prime}}(V) W_{M^{\prime}\!,M}(V)\ - \nn \\
& & P_M(V) \sum_{M^{\prime}} W_{M, M^{\prime}}(V),
\label{eq:master}
\eea
with $W_{M,M^{\prime}}(V)$ the transition rate from atomic spin state $|m_1,..,m_N\ra^{(0)}$ to $|m_1^{\prime},..,m_N^{\prime}\ra^{(0)}$. 
For small tunneling current, i.e. small tip (electrode)-atom coupling, the
atomic chain is approximately in equilibrium and $P_M(V)$ can be approximated by the equilibrium population~\cite{delg10}, which is given by
the stationary solution of the master equation (\ref{eq:master}) (see the derivation in the appendix). Substitution of this solution and Eq.~(\ref{eigenstate-gestoord}) into Eq.~(\ref{eq:I_V}) 
yields the current $I(V)$:
\be
I(V) = G_S\, ( I^{(0)} (V) + I^{(1)}(V) ).
\label{eq:current}
\ee
Here
\bea
I^{(0)}(V) & = & \prod_{i=1}^N \sum_{\stackrel{m_i=-2}{s=\pm}}^2 P_M(V)\, \left\{ (m_j)^2\, F_{0,0,s}(V) \right. \nn \\
& & + \frac{C_{m_j}}{2}\, F_{1,0,s}(V) \nn \\
& & \left. +\, \frac{C_{m_j-1}}{2}\, F_{-1,0,s}(V) \right\}
\label{eq:current0}
\eea
\bea
I^{(1)}(V) & = & \frac{1}{8}\, E^2\, \prod_{i=1}^N \sum_{\stackrel{m_i=-2}{s=\pm}}^2 P_M(V) \nn \\
& & \left\{ 8\, (A_{m_j}^2 + A_{m_j-2}^2)\, F_{0,0,s}(V) \right. \nn \\
& + & C_{m_j}\, B_{m_j,2}^2\, F_{1,0,s}(V) \nn \\ 
& + & 
C_{m_j-1}\, B_{m_j,0}^2\, F_{-1,0,s}(V) \nn \\
& + & A_{m_j-2}^2 \left( C_{m_j-3}\, F_{-3,-2,s}(V) \right. \nn \\
& & \left. \hspace*{1.2cm} +\ C_{m_j-2}\, F_{-1,-2,s}(V) \right) \nn \\
& + & \left.  A_{m_j}^2 \left(
C_{m_j+1}\, F_{1,2,s}(V) + 
C_{m_j+2}\, F_{3,2,s}(V) \right)
\right\}, \nn \\
& & 
\label{eq:currentE}
\eea
with
\bea
A_{m_j+n}^2 & \equiv & A_{m_{j-1},m_j+n,m_{j+1}}^2 
\label{eq:Amj2} \\
B_{m_j,n}^2 & \equiv & A_{m_{j-2},m_{j-1},m_j}^2 + A_{m_{j-2},m_{j-1},m_j-1+n}^2 + \nn \\
& & A_{m_{j-2},m_{j-1}-2,m_j}^2 + A_{m_{j-2},m_{j-1}-2,m_j-1+n}^2 + \nn \\
& & A_{m_{j-1},m_j-1+n,m_{j+1}}^2 + A_{m_{j-1},m_j-3+n,m_{j+1}}^2 + \nn \\
& & A_{m_j,m_{j+1},m_{j+2}}^2 + A_{m_j-1+n,m_{j+1},m_{j+2}}^2 + \nn \\ 
& & A_{m_j,m_{j+1}-2,m_{j+2}}^2 + A_{m_j-1+n,m_{j+1}-2,m_{j+2}}^2  \nn \\
& & 
\label{eq:Bmj2} \\
C_{m_j} & \equiv & 6 - m_j (m_j + 1) 
\label{eq:Cmj} \\
F_{n_1,n_2,s} & \equiv & \frac{eV - s \Del_{n_2,n_1}}{ 1 - e^{- s\beta (eV - s\Del_{n_2,n_1})} } 
\label{eq:Fn1n2} \\
\Del_{n_2,n_1} & \equiv & E_{m_1,..,m_j+n_2,..,m_N} - E_{m_1,..,m_j+n_1,..m_N} \nn \\ 
& = & (n_2 - n_1)\, \left( (2 m_j + n_1 + n_2)D\ + \right. \nn \\ 
& & \left. (m_{j-1} + m_{j+1})J - E_Z \right).
\eea
$I^{(0)}(V)$ is the (zeroth-order) tunneling current in the absence of transverse anisotropy (for $E=0$) and $I^{(1)}(V)$ the additional contribution to this
current for nonzero $E$ (up to first order in perturbation theory).  $P_{M}(V)$ denotes the equilibrium population of state $\psi_{m_1,..,m_N}$ (see also the appendix). 
The coefficient $A_{m_{j-1},m_j,m_{j+1}}$ is given by Eq.~(\ref{eq:Ami}). Fig.~\ref{fig:scheme} schematically illustrates the eigenstates corresponding to
the coefficients $A_{m_j+n}$ and $B_{m_j,n}$ ((\ref{eq:Amj2}) and (\ref{eq:Bmj2})) for $m_j = m_1 = 2$.
\\
\subsection{Transition Rates}
\label{subsec-transitionrates}
In this section we derive expressions for the transition rates $W_{m_1,..,m_N^{\prime}} \equiv W_{m_1,..,m_N,m_1^{\prime},..,m_N^{\prime}}$ between eigenstates 
$|m_1,..,m_N\ra$ and $|m_1^{\prime},..,m_N^{\prime}\ra$ of the atomic spin chain up to first order in the transverse anisotropy energy $E$. 
These rates are also used to calculate the equilibrium occupation $P_M(V)$ of the energy levels, see the appendix.
When an electron tunnels from the STM tip to the surface, or vice versa, and interacts with the atomic spin chain six types of spin transitions can 
occur~\cite{delg10}, denoted by the rates $W_{m_1,..m_N}^{S\to T}(V)$, $W_{m_1,..,m_N^{\prime}}^{S\to S}$, $W_{m_1,..,m_N^{\prime}}^{S \rightarrow T}(V)$,
$W_{m_1,..m_N}^{T\to S}(V)$, $W_{m_1,..,m_N^{\prime}}^{T\to T}$, $W_{m_1,..,m_N^{\prime}}^{T \rightarrow S}(V)$. Because of symmetry, $W_{m_1,..,m_N^{\prime}}^{S\to S}=
W_{m_1,..,m_N^{\prime}}^{T\to T}$, and the pairs of rates $W_{m_1,..m_N}^{S\to T}(V)$, $W_{m_1,..m_N}^{T\to S}(V)$ and $W_{m_1,..,m_N^{\prime}}^{S \rightarrow T}(V)$, 
$W_{m_1,..,m_N^{\prime}}^{T \rightarrow S}(V)$
are identical upon reversal of the bias voltage, i.e. $W_{m_1,..m_N}^{S\to T}(V) = W_{m_1,..m_N}^{T\to S}(-V)$ and $W_{m_1,..,m_N^{\prime}}^{S \rightarrow T}(V) = 
W_{m_1,..,m_N^{\prime}}^{T \rightarrow S}(-V)$. Below we first discuss the physical process described by the rates $W_{m_1,..m_N}^{S\to T}(V)$, 
$W_{m_1,..,m_N^{\prime}}^{S\to S}$, and $W_{m_1,..,m_N^{\prime}}^{S \rightarrow T}(V)$ (which also applies to resp. $W_{m_1,..m_N}^{T\to S}(V)$, 
$W_{m_1,..,m_N^{\prime}}^{T\to T}$, and $W_{m_1,..,m_N^{\prime}}^{T \rightarrow S}(V)$ for reversed bias voltage) and then calculate these rates up to first order in $E$.
\\
1. {\it Elastic tunneling} - $W_{m_1,..m_N}^{S\to T}(V)$ denotes the rate for an electron tunneling from surface (S) to tip (T) without interacting with the atomic spin, i.e. without inducing  a spin transition. 
This rate contributes to the elastic tunneling current. \\
2. {\it Substrate-induced relaxation} - $W_{m_1,..,m_N^{\prime}}^{S\to S}$ corresponds to the simultaneous creation of an electron-hole pair in the surface electrode and a flip of the atomic spin from state 
$|m_1,..,m_N\ra$ to state $|m_1^{\prime},..,m_N^{\prime}\ra$. 
This rate thus does not contribute to the current but does contribute to the equilibrium population $P_M$ at voltages that are sufficiently high for the atomic spin chain to be in an excited state.
At low bias voltages $(W_{m_1,..,m_N^{\prime}}^{S\to S})^{-1}$ is a measure for $T_1$, the atomic 
spin relaxation time~\cite{delg10}. \\
3. {\it Spin-flip assisted inelastic tunneling} - $W_{m_1,..,m_N^{\prime}}^{S \rightarrow T}(V)$ describes the transfer of an electron from surface to tip combined with a transition of 
the spin chain from spin state $|m_1,..,m_N\ra$ to state $|m_1^{\prime},..,m_N^{\prime}\ra$.
This process thus both contributes to the atomic spin dynamics and to the inelastic tunneling current.
For an unpolarized STM tip the three rates can be calculated to lowest order in the electrode-chain coupling using Fermi's golden rule. This results in:
\bea
W_{m_1,..m_N}^{{\rm S}\, \to \, {\rm T}}(V)
& = & \frac{4\pi}{\hbar} W_1\,  \frac{eV}{1 - e^{-\beta eV}} \times \nn \\
& & \left| \sum_{\al=0} \la m_1,..m_N | S_{\al}^{(j)} | m_1,..m_N \ra \right|^2 \nn \\
& = & \frac{4\pi}{\hbar} W_1 \frac{eV}{1 - e^{-\beta e V}}
\label{eq:WW1}\\
W_{m_1,..m_N^{\prime}}^{{\rm S}\, \to \, {\rm S}}
& = & \frac{4\pi}{\hbar} W_2 \frac{\Del_{M,M^{\prime}}}{1 - e^{-\beta \Del_{M,M^{\prime}}}}\ \times \nn \\
& & \left| \sum_{\al=x,y} \la m_1,..m_N | S_{\al}^{(j)} | m_1^{\prime},..m_N^{\prime} \ra \right|^2 \nn \\
\label{eq:WW2}\\
W_{m_1,..,m_N^{\prime}}^{{\rm S}\, \to \, {\rm T}}(V)
& = & \frac{4\pi}{\hbar} W_3\, \frac{eV + \Del_{M,M^{\prime}}}{1 - e^{-\beta (eV + \Del_{M,M^{\prime}})}}\, \times \nn \\
& & \left| \sum_{\al=x,y,z} \la m_1,..m_N | S_{\al}^{(j)} | m_1^{\prime},..m_N^{\prime} \ra \right|^2, \nn \\
\label{eq:WW3}
\eea
with $W_1 \equiv \rho_S \rho_T T_0^2$, $W_2 \equiv \rho_S^2 T_J^2$ and $W_3 \equiv \rho_S \rho_T T_J^2$. 
$T_0$ and $T_J$ correspond to the direct (spin-independent) tunnel coupling and the tunneling-induced exchange coupling, respectively~\cite{delg10,zhan13}. 
Calculating the total spin transition rates $W_{m_1,..m_N}^{{\rm S}\, \to \, {\rm S}}$ $\equiv$  $\prod_{i=1}^N \, \left( \sum_{m_i^{\prime}=-2}^{2} \right) \, W_{m_1,..,m_N^{\prime}}^{{\rm S}\, \to \, {\rm S}}$ and 
$W_{m_1,..,m_N}^{{\rm S}\, \to \, {\rm T}}(V)$ $\equiv$ $\prod_{i=1}^N \, \left( \sum_{m_i^{\prime}=-2}^{2}\right) \, W_{m_1,..,m_N^{\prime}}^{{\rm S}\, \to \, {\rm T}}(V)$ for the eigenstates $\psi_{m_1,..m_N}$ [Eq.~(\ref{eigenstate-gestoord})]
we obtain, up to first order in $E$ and for the STM tip coupled to atom $j$,
\bea
W_{m_1..m_N}^{{\rm S}\, \to \, {\rm S}} & = & 
\frac{2\pi}{\hbar} W_2\, \left\{ C_{m_j} F_{1,0,+}(0)\, + C_{m_j-1} F_{-1,0,+}(0)  \right. \nn \\
& + &  \frac{E^2}{4}\, \left( 
C_{m_j} B_{m_j,2}^2\, F_{1,0,+}(0) \right. \nn \\
& + & C_{m_j-1} B_{m_j,0}^2\, F_{-1,0,+}(0) \nn \\
& + &A_{m_j-2}^2\, (C_{m_j-3} F_{-3,-2,+}(0) \nn \\
& + &
C_{m_j-2} F_{-1,-2,+}(0))  \nn \\
& + & \left. \left. 
A_{m_j}^2\, ( C_{m_j+1} F_{1,2,+}(0) + 
C_{m_j+2} F_{3,2,+}(0) ) \right)  \right\} \nn \\
& & 
\label{eq:matrixelement6}
\eea
\bea
W_{m_1..m_N}^{{\rm S}\, \to \, {\rm T}}(V) & = & 
\frac{2\pi}{\hbar} W_3\, \left\{ 2 m_j^2 F_{0,0,+}(V) + \right. \nn \\ & & 
C_{m_j} F_{1,0,+}(V)\, +\, C_{m_j-1} F_{-1,0,+}(V) + \nn \\ 
& & \frac{E^2}{4}\, \left( 
8\, (A_{m_j}^2 + A_{m_j-2}^2)\, F_{0,0,+}(V) +
\right. \nn \\
& & C_{m_j} B_{m_j,2}^2\, F_{1,0,+}(V) + \nn \\
& & C_{m_j-1} B_{m_j,0}^2\, F_{-1,0,+}(V) + \nn \\
& & A_{m_j-2}^2\, ( C_{m_j-3} F_{-3,-2,+}(V) + \nn \\
& & C_{m_j-2} F_{-1,-2,+}(V)) + \nn \\
& & 
A_{m_j}^2\, ( C_{m_j+1} F_{1,2,+}(V) + \nn \\
& & \left. \left. C_{m_j+2} F_{3,2,+}(V) ) \right)
\right\}. \nn \\
& & 
\label{eq:matrixelement5}
\eea
Here $A_{m_j+n}^2$, $B_{m_j,n}^2$, $C_{m_j}$ and $F_{n_1,n_2,s}$ 
are given by Eqns.~(\ref{eq:Amj2})-(\ref{eq:Fn1n2}).
\section{$I(V)$, $dI/dV$ and $d^2I/dV^2$}
\label{sec-current}
We now calculate the inelastic tunneling current [Eq.~(\ref{eq:current})], the corresponding differential conductance and the IETS spectra for 
the ground state of a chain consisting of $N$ atoms with antiferromagnetic coupling and analyze
the effect of the transverse anisotropy energy $E$. For the (N\'eel-like) ground state $\psi_{-2,2,..,-2,2}$ [Eq.~({\ref{eigenstate-gestoord}) with $m_i=-2$ for $i$ odd and
$m_i=2$ for $i$ even] and the STM tip located above the first atom we
obtain: 
\be
I_{\text{N\'eel}}(V) =  G_S\, \left( I^{(0)}_{\text{N\'eel}}(V) + I^{(1)}_{\text{N\'eel}}(V) \right) 
\label{eq:currentNeel}
\ee
with 
\bea
I^{(0)}_{\text{N\'eel}}(V) & = & 2\, \sum_{s=\pm} \left( 2\, F_{0,0,s}(V) + F_{1,0,s}(V) \right) 
\label{eq:current0Neel} \\
I^{(1)}_{\text{N\'eel}}(V) & = & \frac{1}{4}\, E^2\, \sum_{s=\pm} \left\{ 4 A_{0,-2,2}^2 F_{0,0,s}(V) \right. \nn \\
&& +\ 2\ B_{-2,2}^2 F_{1,0,s}(V) \nn \\ 
& & \left. +\, 3\, A_{0,-2,2}^2 (F_{1,2,s}(V) + F_{3,2,s}(V)) \right\}.
\label{eq:currentENeel}
\eea
\begin{figure}[h]
\includegraphics[width=0.5\textwidth]{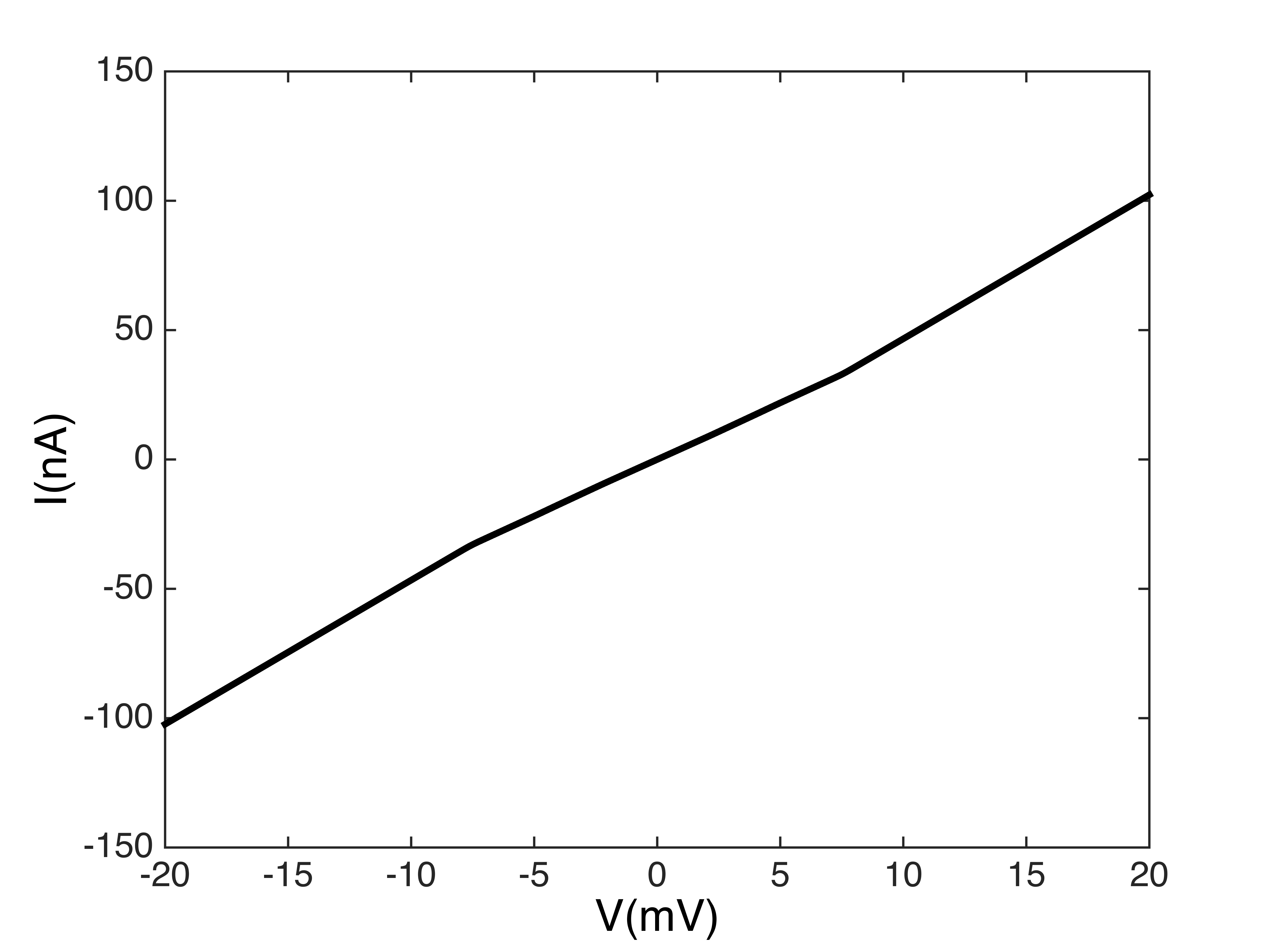}
\caption{Inelastic tunneling current $I$ [Eq.~(\ref{eq:current})] normalized to the zero-bias conductance $G_S$ for an antiferromagnetic chain consisting of four atoms in the ground state 
$\psi_{2,-2,2,-2}$ with the STM tip coupled to the first atom. 
Parameters used are $D=-1.3$ meV, $E=0.3$ meV, $J=1.7$ meV, and $T=1$ K.}
\label{fig:IVAFM}
\end{figure}
\begin{figure}[h]
\includegraphics[width=0.5\textwidth]{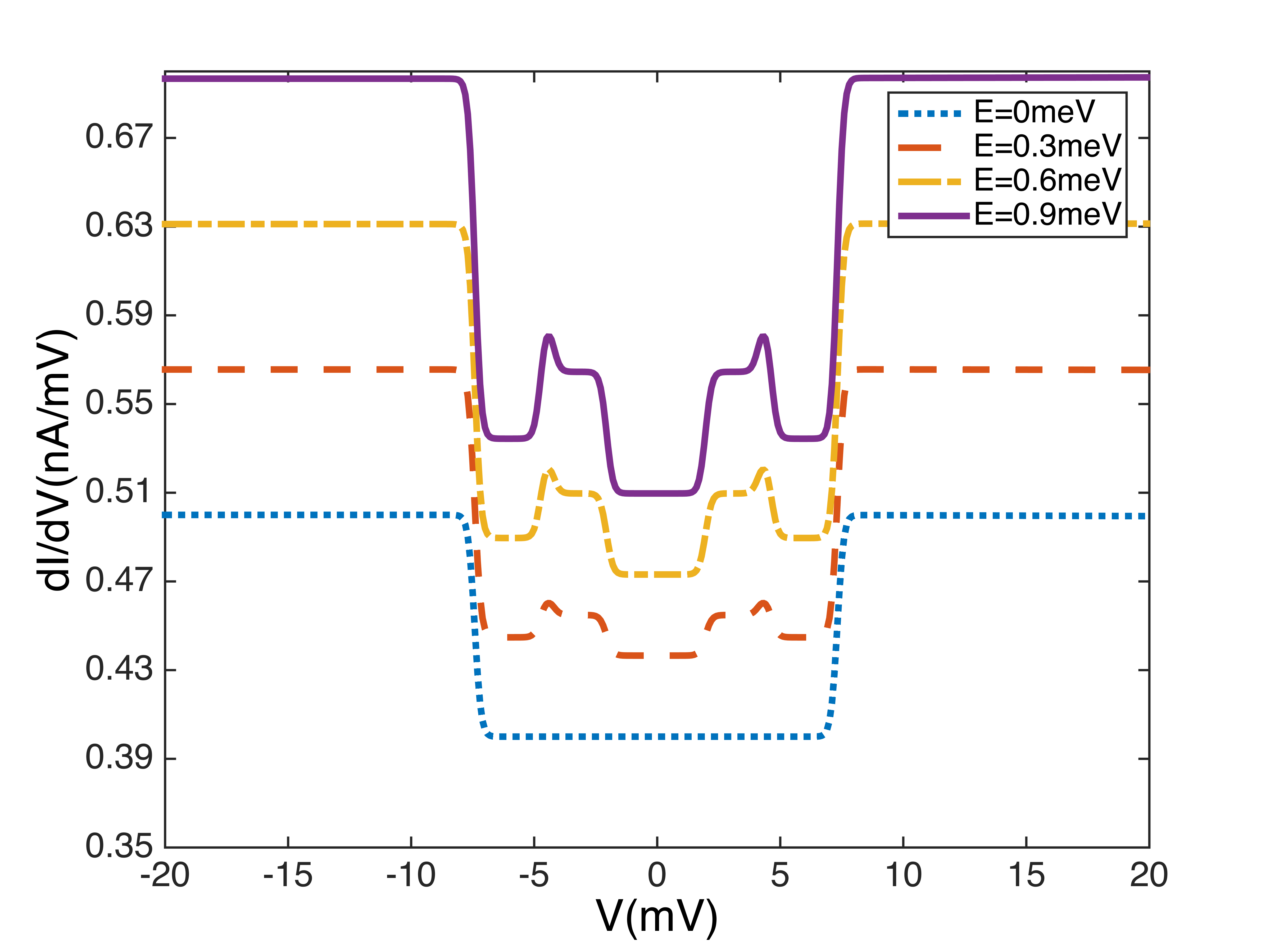}
\caption{(color online) $dI/dV$ normalized to the zero-bias conductance $G_S$ for the chain in Fig.~\ref{fig:IVAFM}.
Curves are shifted for clarity.}
\label{fig:TOTAFM}
\end{figure}
\begin{figure*}[ht]
\centering
\begin{minipage}{0.19\textwidth}
  \centering
  \includegraphics[width=1.\linewidth]{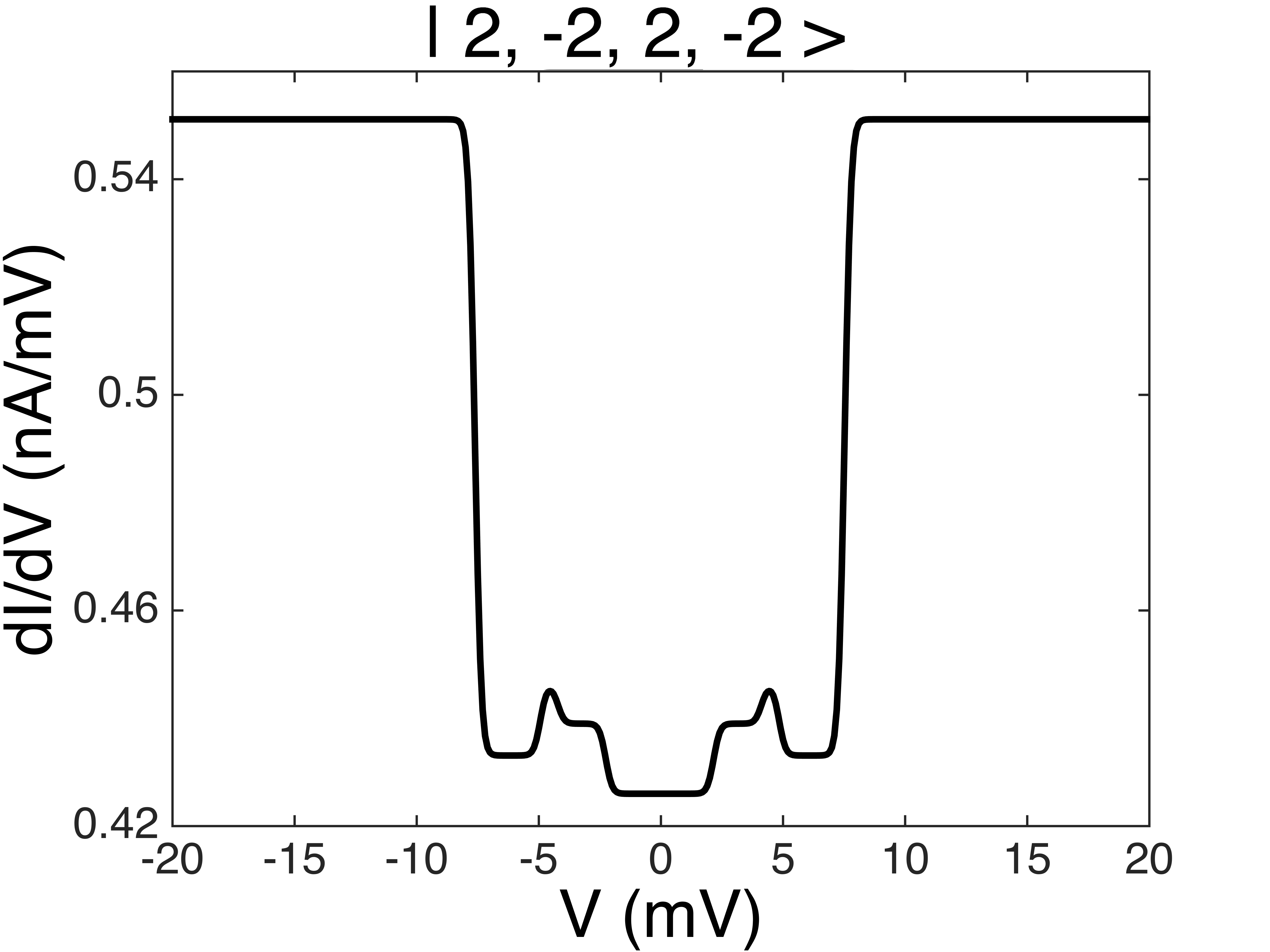}
  \label{fig:test1}
\end{minipage}%
\begin{minipage}{.19\textwidth}
  \centering
  \includegraphics[width=1.\linewidth]{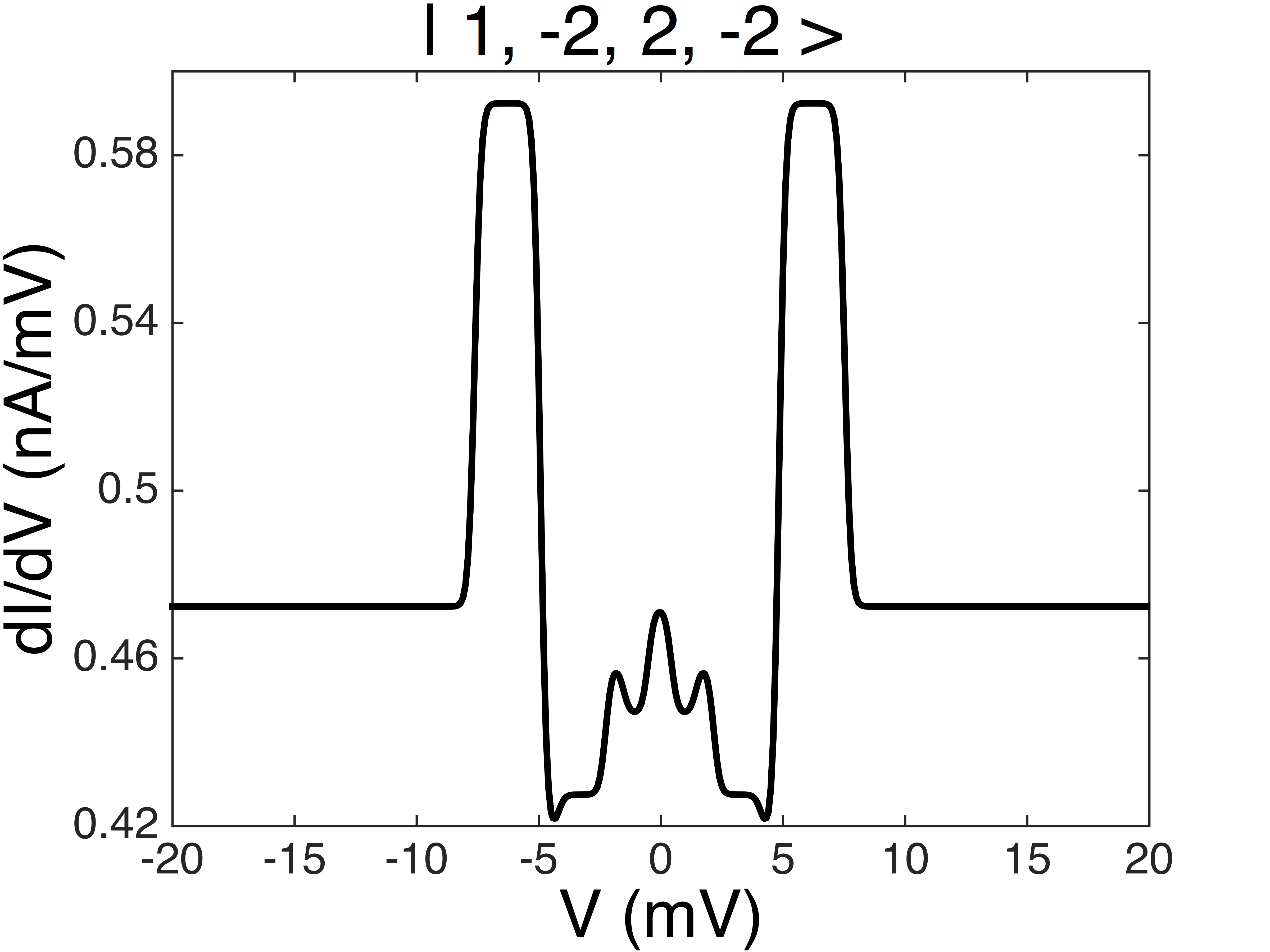}
   \label{fig:test2}
\end{minipage}
\begin{minipage}{.19\textwidth}
  \centering
  \includegraphics[width=1.\linewidth]{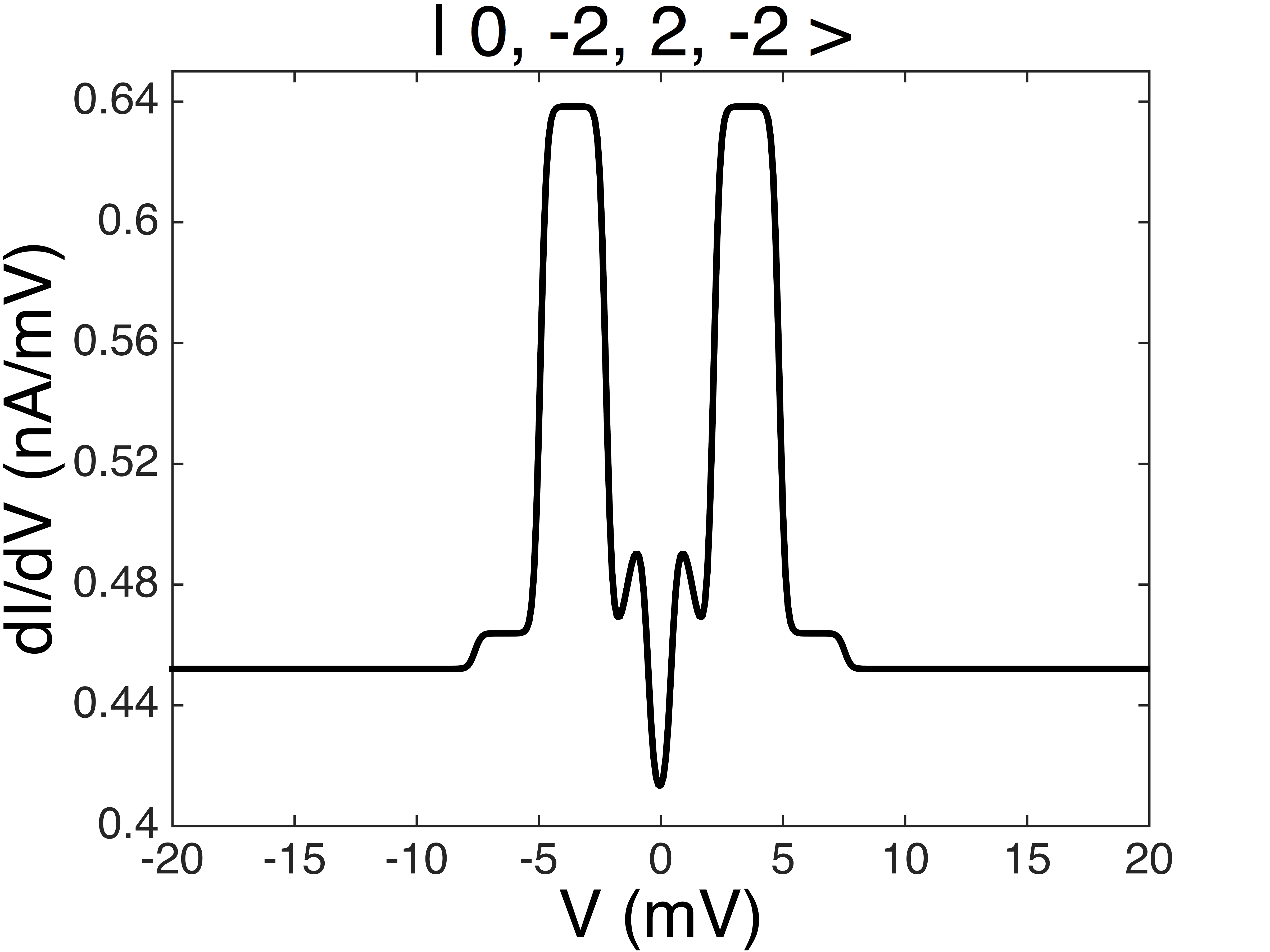}
  \label{fig:test3}
\end{minipage}
\begin{minipage}{.19\textwidth}
  \centering
  \includegraphics[width=1.\linewidth]{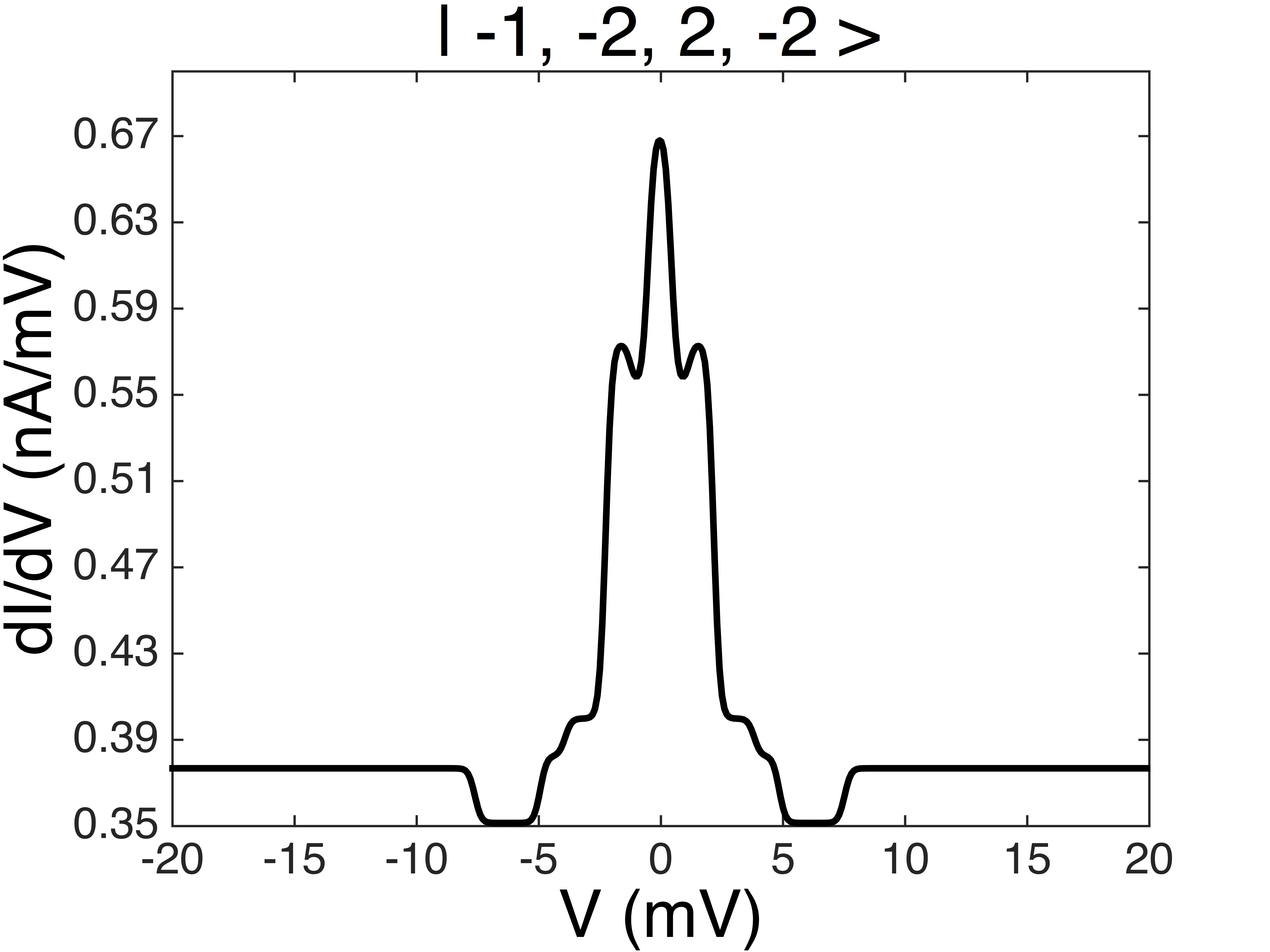}
  \label{fig:test3}
\end{minipage}
\begin{minipage}{.19\textwidth}
  \centering
  \includegraphics[width=1.\linewidth]{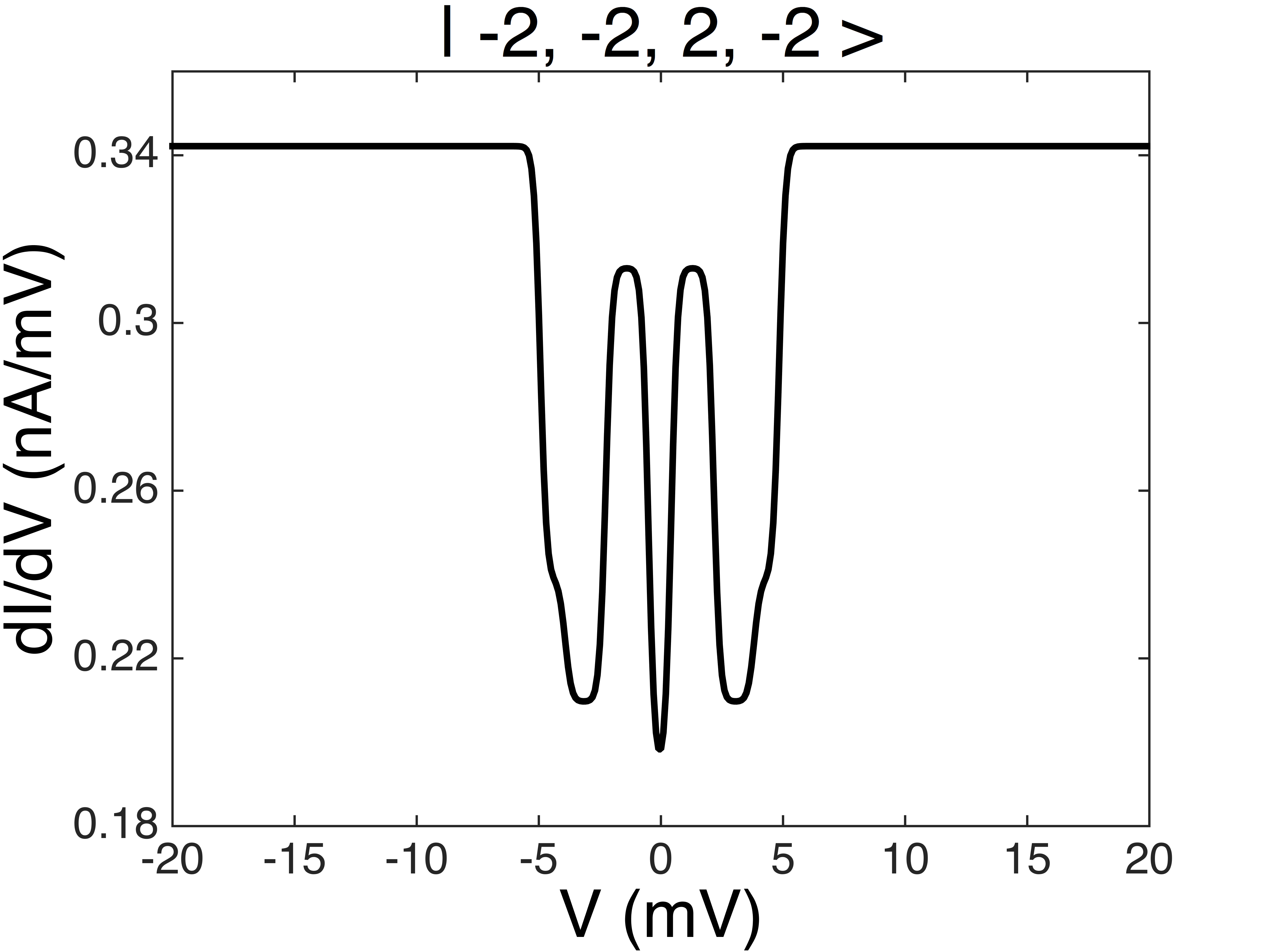}
  \label{fig:test3}
\end{minipage}
\caption{$dI/dV$ (normalized to the zero-bias conductance) of the five separate states $|m_1,-2,2,-2\ra$ with $m_1 \in [2,..,-2]$ in the sums of Eqns.~(\ref{eq:current0}) and (\ref{eq:currentE}) for an antiferromagnetic chain in the
 ground state $\psi_{2,-2,2,-2}$ with the STM tip coupled to the first atom. Parameters used are the same as in Fig.~\ref{fig:IVAFM}.}
\label{fig:DIDVAFM}
\end{figure*}
The tunneling currents (\ref{eq:current0Neel}) and (\ref{eq:currentENeel}) depend on three energy gaps, corresponding to transitions between atomic spin levels with different values 
of $m_1$:
\bea
\Del_1 & \equiv & \Del_{0,1} = 3D - 2J + E_z \ \ {\small (m_1\! =\! -2\! \rightarrow\! m_1\! =\! -1)} \nn \\
\Del_2 & \equiv &  \Del_{2,1} = - D + 2J - E_z  \ \ (m_1\! =\! 0 \rightarrow m_1=-1) \nn \\
\Del_3 & \equiv &  \Del_{2,3} = - D - 2J + E_z \ \ (m_1=0 \rightarrow m_1=1). \nn \\
\label{eq:Neel2}
\eea
For the other N\'eel-like ground state $\psi_{2,-2,..,2,-2}$ the expressions (\ref{eq:currentNeel})-(\ref{eq:Neel2}) are equivalent with only the signs of $m_i$, $i=1,..,N$, and the sign of $E_z$
reversed. Since $E_z \ll D, |J|$ the energy gaps (\ref{eq:Neel2}) are practically the same in both cases. 
In the derivation of Eq.~(\ref{eq:currentNeel}) we have taken $P_M=1$ for the ground state and zero otherwise, since at low temperatures and voltages ($k_BT, eV \ll |\Del_1|$) 
the equilibrium population of the excited states is negligible (see also Fig.~\ref{fig:PM} and discussion thereof
in the text). The assumption that $P_M(V)$ is given by the equilibrium population may not be valid anymore at higher voltages when non-equilibrium effects start to play a role~\cite{2delg10}.
Experimentally, $P_{|-2,2,..,-2,2\ra^{(0)}}=1$ corresponds to using e.g. a half-metal tip.
\\
\noi Inspection of Eq.~(\ref{eq:currentNeel}) using the requirement $I^{(1)}(V) \ll I^{(0)}(V)$ shows that our perturbative approach is valid for values
of transverse anisotropy $E^2 \ll J^2, (D-J)^2$. 
\\
\noi Fig.~\ref{fig:IVAFM} shows the current $I(V)$ [Eq.~(\ref{eq:current}), including all spin states and equilibrium populations $P_M$ for each state $|m_1,..,m_N\ra^{(0)}$] for typical experimental values~\cite{khat13,spin14,yan15} of the transverse anisotropy strength $E$. 
As expected, the current increases linearly with $V$. It shows a kink (change of slope) at $eV \approx \pm |\Del_1| \approx \pm 7.5$\, meV, which corresponds to the energy gap
between the ground state $|2,-2,..,2,-2\ra^{(0)}$ and the first excited state $|1,-2,..,2,-2\ra^{(0)}$ for $E=0$ of the atomic spin chain (the same argument applies for the other ground state
$|-2,2,..,-2,2\ra^{(0)}$). The increase in slope is to a very good approximation given by the coefficients of the corresponding 
activation energy terms $F_{1,0,s}(V)$ in Eqns.~(\ref{eq:current0Neel}) and (\ref{eq:currentENeel}). The finite transverse anisotropy energy $E$ introduces additional kinks in the voltage 
region $-|\Del_1| < eV < |\Del_1|$. This can be seen more
clearly in Fig.~\ref{fig:TOTAFM}, which shows the differential conductance $dI/dV$ for the same chain for several values of the transverse anisotropy energy $E$. The 
large stepwise increase in $dI/dV$ at $eV \approx \pm |\Del_1| \approx \pm 7.5$ meV in the figure corresponds to the kinks at these energies in Fig.~\ref{fig:IVAFM}. In addition, 
however, also steps in $dI/dV$ occur at voltages $eV \approx \pm |\Del_2|$ and $eV \approx \pm |\Del_3|$. These correspond to transitions between higher-lying excited states: 
The step in $dI/dV$ at $eV \approx |\Del_3| \approx 2.3$\, meV corresponds to the excitation 
from spin state $|0,-2,..,2,-2\ra^{(0)}$ to state $|-\!1,-2,..,2,-2\ra^{(0)}$. Then,
around $eV \approx 4.9$\, meV, a second step occurs, corresponding to the excitation from state $|1,-2,..,2,-2\ra^{(0)}$ to $|0,-2,..,2,-2\ra^{(0)}$. At this energy the excited state 
$|1,-2,..,2,-2\ra^{(0)}$ has become somewhat populated allowing for this transition to occur~\cite{step}. At slightly higher voltage, however, a steplike decrease occurs, corresponding
to decay from state $|0,-2,..,2,-2\ra^{(0)}$ to state $|1,-2,..,2,-2\ra^{(0)}$. Here the spin chain thus undergoes a transition from a higher-lying to a lower-lying state
and an electron tunnels from drain (the STM tip) to source (the surface), thereby lowering the rate of increase of $I(V)$. Decays in differential conductance have been 
experimentally observed in STM measurements of magnetic atoms (see e.g. Ref.~\cite{otte08,loth10}) and were explained in terms of the non-equilibrium occupation of 
different spin levels, i.e. competition between depletion of one spin level in favour of another multiplied by the intensity (determined by the matrix elements) of the corresponding transitions. 
\\
\noi Fig.~\ref{fig:DIDVAFM} provides a more detailed illustration of the competition between these two processes. This figure shows each of the five terms $m_1$ $\in$ $[2,...,-2]$\ that
contribute to $dI/dV$ in Fig.~\ref{fig:TOTAFM} separately (the current (\ref{eq:current}) is the sum of these five terms weighed by the equilibrium population $P_M$ for each state). 
When inspecting the figure, we see that in the panels corresponding to $m_1=1$ and $m_1=0$
a sharp increase of $dI/dV$ occurs at, resp., energies $eV \approx |\Del_2| \approx 4.9$ meV and $eV \approx |\Del_3| \approx 2.3$ meV. Here the chain undergoes a transition to the next higher-lying state
(from $m_1=1$ to $m_1=0$ and from $m_1=0$ to $m_1=-1$, resp.) In the same two panels the differential conductance subsequently decreases at voltages $eV \approx |\Del_1| \approx 7.5$ meV 
and $eV \approx |\Del_2| \approx4.9$ meV, 
when the spin chain decays to the next lower-lying state. Similar analysis applies for the steps in the other panels. 
\\
Note that the position at which steps in $dI/dV$ occur does not depend on the strength of the transverse anisotropy, since the energy gaps $\Del_1 - \Del_3$ in Eq.~(\ref{eq:Neel2}) are 
independent of $E$ (up to first order in $E$). The onset of these in-gap steps is affected when $D$ is not constant along the chain, but varies from atom to atom (as e.g. in the experiment 
described in Ref.~\cite{yan15}). Within the parameter range of $D$, $J$ and $E$ that we consider (described in Sec.~\ref{sec-conclusion}) atom-to-atom variations in $D$ 
lead to small variations in the onset and the heights of the steps. The latter step heights at $eV=\Del_{n_1,n_2}$ scale with $E^2$ and are
to a good approximation given by the (sum of the) prefactors $F_{n_1,n_2,s}$ in Eq.~(\ref{eq:currentNeel}); for example, the step height at $eV \approx \pm |\Del_3|$ is 
given by $(3/4) E^2 A_{0,-2,2}^2 \approx E^2/(D-J)^2$ (for small
Zeeman energy). This step height is a direct measure for the spin excitation transition intensity. Finally, in the limit $|eV| \gg |\Del_1|$, the differential conductance saturates at 
\bea
\frac{dI_{\text{N\'eel}}}{dV} & \stackrel{|eV| \gg |\Del_1|}{\rightarrow} & 6 e G_S \left(1 + \frac{E^2}{12}\, (5 A_{0,-2,2}^2 + 2 B_{-2,2}^2) \right) \nn \\
& \approx & 6 e G_S \left(1 + \frac{E^2}{8}\, \left( \frac{5}{(D-J)^2} + \frac{3}{J^2} \right) \right). \nn \\
& & \label{eq:limitingvalue}
\eea
The second line in Eq.~(\ref{eq:limitingvalue}) is valid when the Zeeman energy is small, $|E_z| \ll |J|, |D-J|$. 
Fig.~\ref{fig:D2IDV2AFM} shows the IETS spectra $d^2I/dV^2$ corresponding to the differential conductance $dI/dV$ in Fig.~\ref{fig:TOTAFM}. The additional peaks 
and valleys induced 
by the finite transverse anisotropy strength in the voltage region between -7.5 meV and 7.5 meV can clearly be seen. \\
Finally, from an experimental point of view Eqns.~(\ref{eq:Neel2}) can be used to extract the values of $D$ and $J$ from $dI/dV$ data. The height of measured in-gap 
steps can subsequently be used to obtain the strength of transverse anisotropy $E$.
\begin{figure}[ht]
\includegraphics[width=0.5\textwidth]{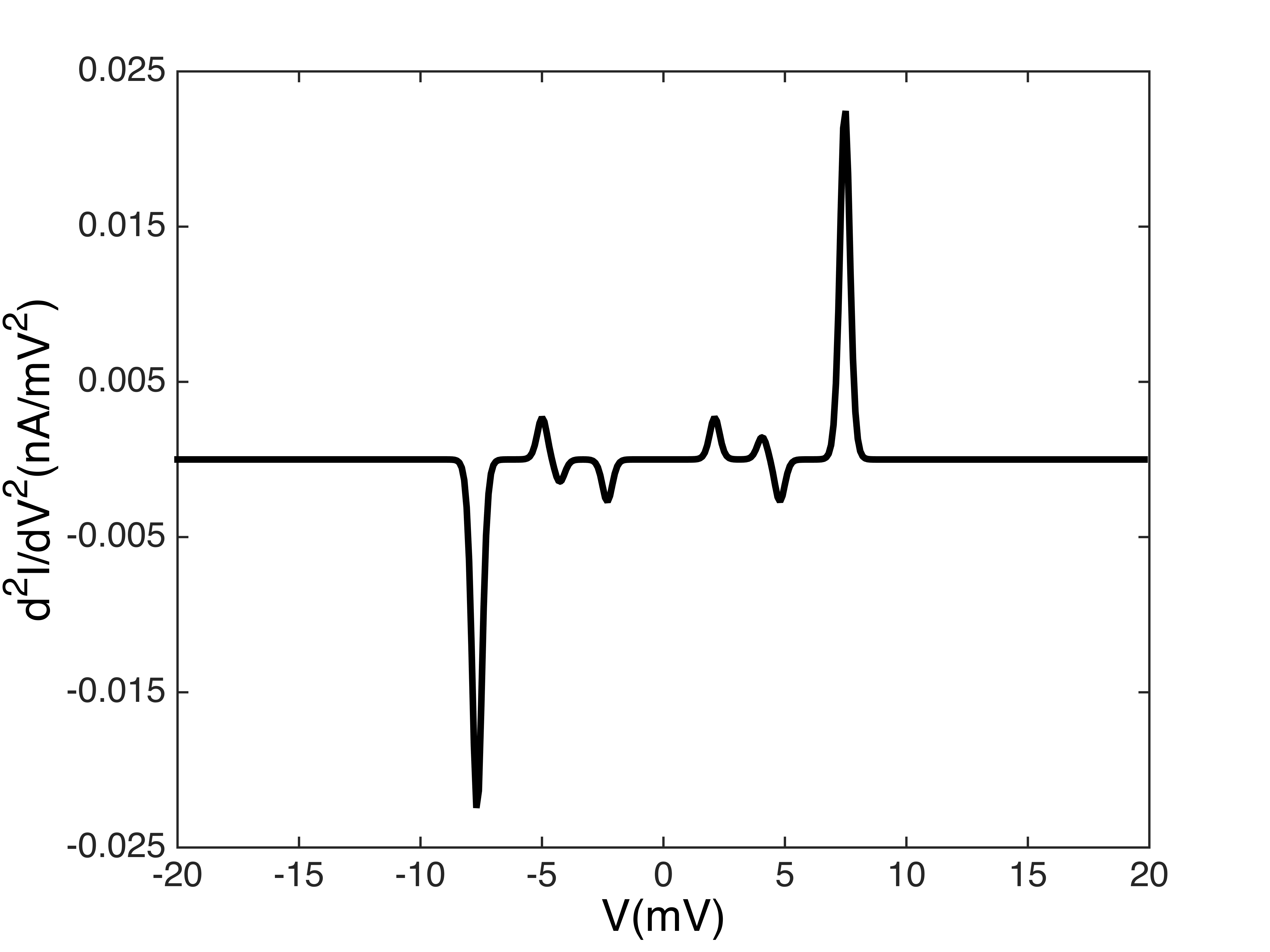}
\caption{$d^2I/dV^2$ spectra corresponding to the differential conductance $dI/dV$ shown in Fig.~\ref{fig:TOTAFM}
for $E=0.3$\, meV.}
\label{fig:D2IDV2AFM}
\end{figure}
\section{Transition Rates}
\noi In this section we analyze the transition rates $W_{m_1,..,m_N}^{S\to S}$ and $W_{m_1,..,m_N}^{S\to T}(V)$ [Eqns.~(\ref{eq:matrixelement6}) and (\ref{eq:matrixelement5})] for the ground state $\psi_{-2,2,..,-2,2}$ 
of an antiferromagnetic $N$-atomic spin chain. By evaluating the matrix element in Eqns.~(\ref{eq:matrixelement6}) and (\ref{eq:matrixelement5}) this results in, for the STM tip coupled to the first atom,
\bea
W_{\text{N\'eel}}^{{\rm S}\, \to \, {\rm S}} & = & 
\frac{8\pi}{\hbar} W_2\, \left\{ F_{1,0,+}(0) + \frac{E^2}{8} \left( 2 B_{-2,2}^2\, F_{1,0,+}(0) \right. \right. \nn \\ 
& & \left. \left. +\, 3 A_{0,-2,2}^2 
(F_{1,2,+}(0) + F_{3,2,+}(0) \right) \right\} 
\label{eq:Neel4}
\eea
and
\bea
W_{\text{N\'eel}}^{{\rm S}\, \to \, {\rm T}}(V) & = & 
\frac{8\pi}{\hbar} W_3\, \left\{ 2\, F_{0,0,+}(V)\ + F_{1,0,+}(V)\, + \right. \nn \\ 
& & \frac{E^2}{8} \left( 4 A_{0,-2,2}^2 F_{0,0,+}(V) \right. \nn \\
& & \ +\ 2 B_{-2,2}^2\, F_{1,0,+}(V) \nn \\
& & \left. \left. +\ 3 A_{0,-2,2}^2 
(F_{1,2,+}(V) + F_{3,2,+}(V)) \right) \right\}, \nn \\
& & 
\label{eq:Neel3}
\eea
with $\Del_1$, $\Del_2$ and $\Del_3$ given by Eqns.~(\ref{eq:Neel2}). 
\begin{figure}[h]
\centering
\begin{minipage}[b]{0.47\textwidth}
\includegraphics[width=1\textwidth]{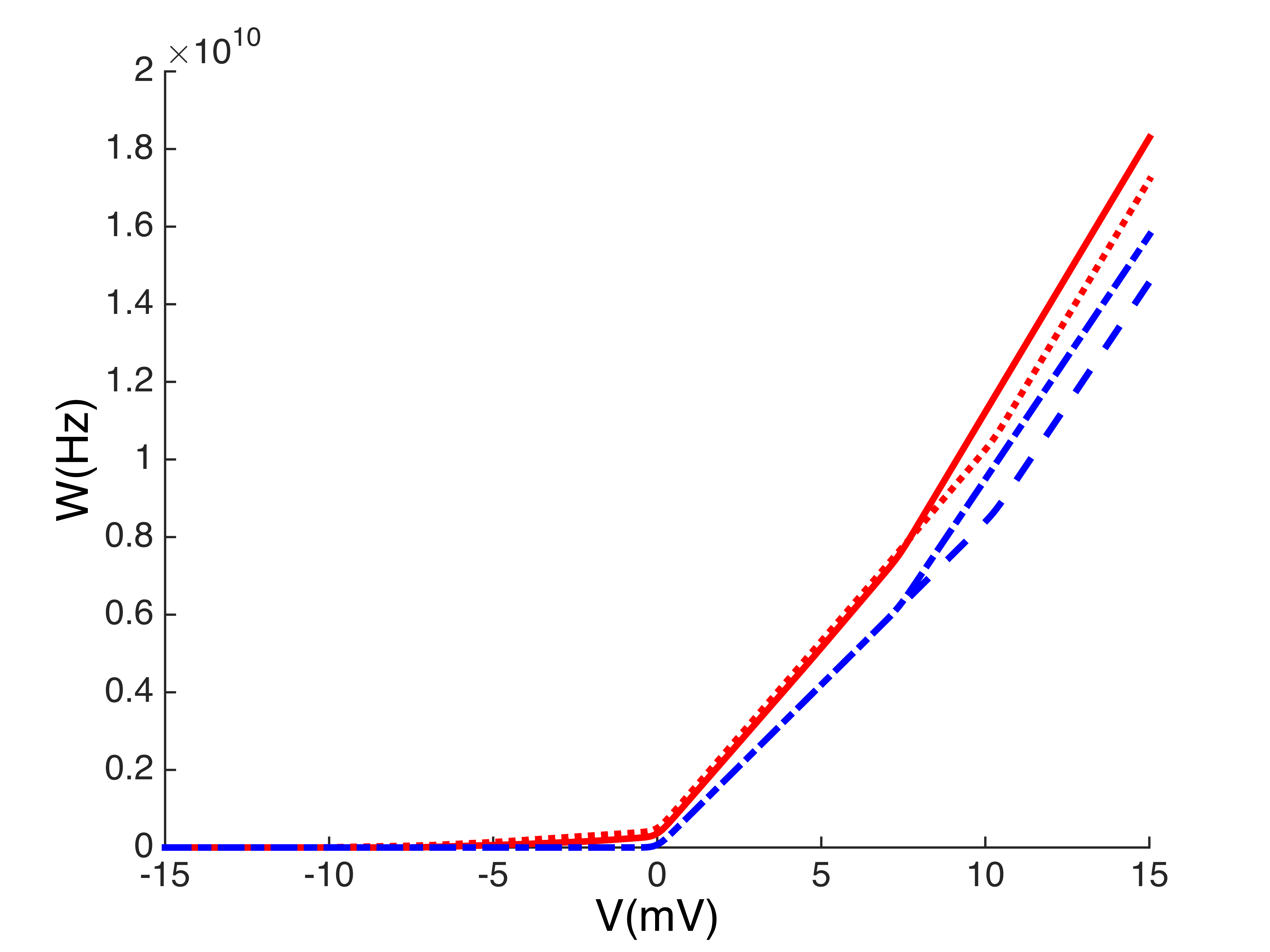}
\caption{(Color online) The spin transition rate $W\equiv W_{\text{N\'eel}}^{S \rightarrow T}(V)$ [Eq.~(\ref{eq:matrixelement5})] for an antiferromagnetic atomic chain 
in the ground state with the STM tip coupled to the {\it first} atom and $E=0.3$ meV or $E=0.6$ meV (blue dot-dashed line and red solid line, respectively) and with the tip coupled to the {\it second} atom 
and $E=0.3$ meV or $E=0.6$ meV (blue dashed line and red dotted line, respectively). Other parameters used are $D=-1.3$\, meV, $J=1.7$\, meV, $B=1$\, T and $W_3=1.1 \times 10^{-5}$ (the latter value is taken from the experiment in Ref.~\cite{spin14}).}
\label{fig:Rates}
\end{minipage}
\hfill
\centering
\begin{minipage}[b]{0.47\textwidth}
\includegraphics[width=1\textwidth]{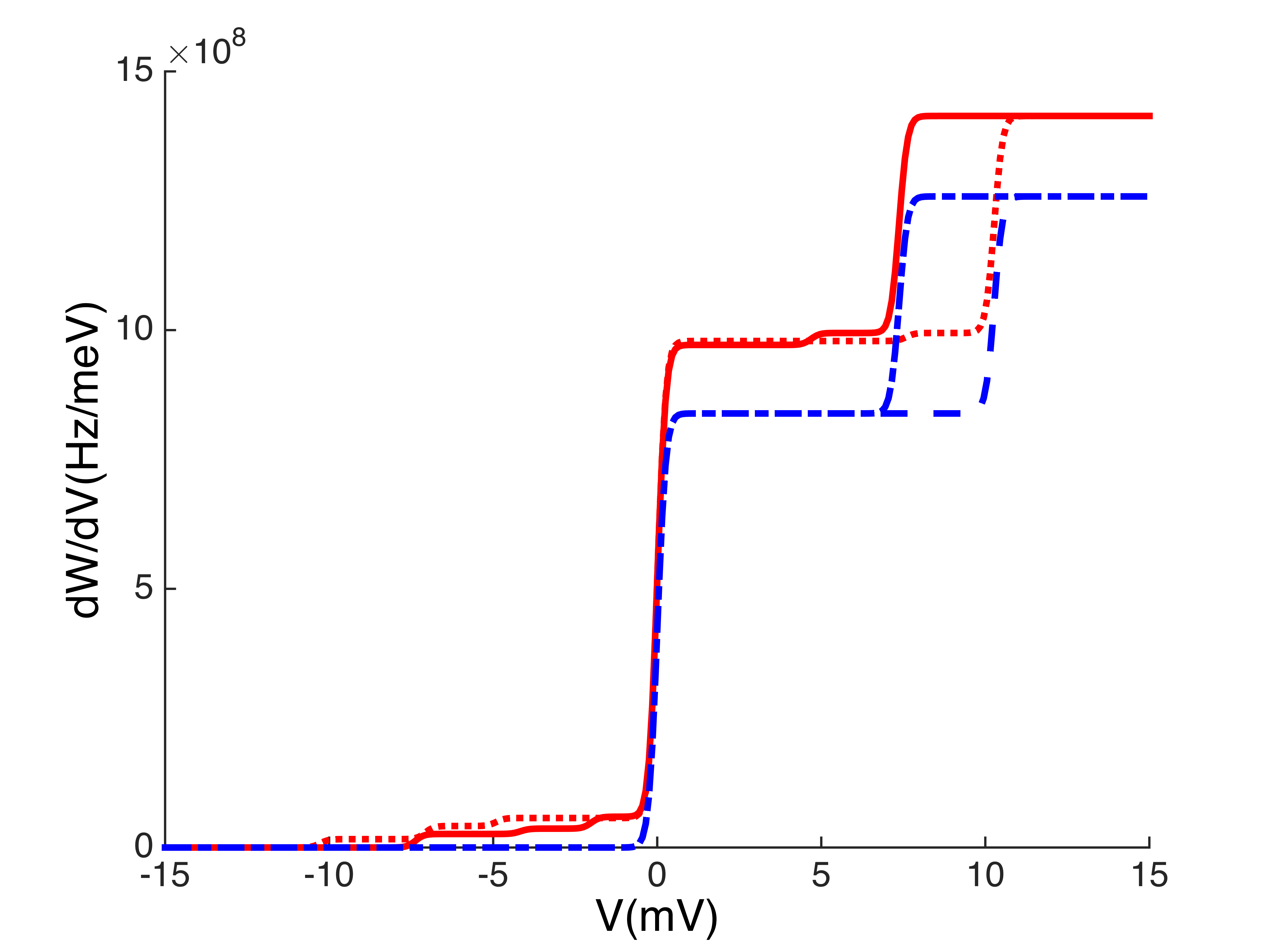}
\caption{(Color online) First derivative of the transition rates $W_{\text{N\'eel}}^{S \rightarrow T}(V)$ in Fig.~\ref{fig:Rates}.}
\label{fig:RatesDWDV}
\end{minipage}
\end{figure}
Fig.~\ref{fig:Rates} shows $W^{S\to T}_{\text{N\'eel}}(V)$ for the STM tip coupled to either the first or the second atom along the chain. As expected,
when the tip interacts with the first atom, $W^{S\to T}_{\text{N\'eel}}(V)$ exhibits a clearly visible kink (change of slope) at the same voltage $eV=|\Del_1| \sim 7.5$ meV 
as the inelastic current in Fig.~\ref{fig:IVAFM}, which corresponds to the energy gap between the ground state and the first excited state of the chain
(for the tip coupled to the second atom this gap is larger, given by $3D-4J-E_Z \approx 10.5$\, meV).
In addition, the finite transverse anisotropy energy also here induces additional kinks at $eV = |\Del_2|$
and $eV = |\Del_3|$. The positions of these kinks can be seen more clearly in the graph of the derivative $dW^{S\to T}_{\text{N\'eel}}/dV$ in Fig.~\ref{fig:RatesDWDV}.
The onset of $W^{S\to T}_{\text{N\'eel}}$ at $V=0$\,meV is due to thermally activated elastic tunneling.
\\
Fig.~\ref{fig:Rates} also shows that finite transverse anisotropy energy increases the spin transition rates $W^{S\to T}_{\text{N\'eel}}(V)$ for any value of the voltage $V$. 
From Eq.~(\ref{eq:Neel3}) and the tip coupled to the first atom we find that this relative increase scales as 
$E^2/(D-J)^2$ for energies $eV \ll |\Del_1|$ and as $\frac{E^2}{8}\, \left( \frac{5}{(D-J)^2} + \frac{3}{J^2} \right)$ for energies $eV \gtrsim |\Del_1|$. 
For the voltage-independent relaxation rate $W^{S\to S}_{\text{N\'eel}}$ (not
shown in Fig.~\ref{fig:Rates}) we obtain from
Eq.~(\ref{eq:Neel4}) for $|E_z| \ll |D-J|, |J|$
\be
W^{S\to S}_{\text{N\'eel}} \approx \frac{8\pi}{\hbar} W_2 \left| \Del_1 \right| \left( 1 + \frac{9}{16}\, \frac{E^2}{J^2} \right).
\nn
\ee
Since $T_1 \sim 1/W^{S\to S}_{\text{N\'eel}}$ at low bias voltages, the presence of finite transverse anisotropy energy thus leads to a decrease of the spin relaxation time which scales as $E^2/J^2$.
\section{Summary and Conclusions}
\label{sec-conclusion}
We have presented a perturbative theory for the effect of single-spin transverse magnetic anisotropy
on tunneling-induced spin transitions in atomic chains with Ising exchange coupling. We qualitatively predict the dependence of the inelastic tunneling current $I$
and the transition rates between atomic spin levels on the transverse anisotropy energy $E$ and show that the presence of finite values of $E$ leads to additional steps in the differential conductance 
$dI/dV$ and to higher spin transition rates. For an antiferromagnetically coupled chain in the N\'eel ground state both the heights of the additional steps and the increase in spin 
transition rates at low bias voltage scale as $E^2/(D-J)^2$, while the latter crosses over to $E^2/J^2$ scaling for higher voltages $eV\gtrsim |\Del_1|$.

Our model is relevant for materials in which the easy-axis exchange interaction dominates over the transverse exchange interaction (justifying the use of the Ising Hamiltonian),
measurements at low current with a non-magnetic STM tip and for values of transverse anisotropy $E^2 /J^2, E^2/(D-J)^2 \ll 1$. The latter requirement is in agreement with typical values 
of $E$, $D$ and $J$ measured in chains 
of, for example, Fe or Mn atoms~\cite{loth12,hein04,hirj06,hirj07,otte08,spin14,khat13,yan15}, where $D$ varies between -2.1 and -1.3 meV, J is in the range 1.15-1.6 meV, and E is 0.3-0.31 meV. 
We therefore expect our results to be applicable for antiferromagnetically coupled 
chains consisting of these and similar magnetic atoms with little-to-none local distortion between atoms and deposited on a flat symmetric substrate, so that spin-orbit interaction (and 
thereby induced Dzyaloshinskii-Moriya interaction) is weak and can be neglected. 

Our model could be used to extract the values of $D$ and $J$ from the onset of the in-gap steps in measurements of $dI/dV$, since these threshold voltages are directly related to energy 
gaps between spin levels, see Eq.~(\ref{eq:Neel2}). The height of these in-gap steps can subsequently be used to extract the value of $E$. In fact, with a note of caution, in-gap features 
in $dI/dV$ may actually be present in Fig.~2(c) of Ref.~\cite{yan15}. Using the parameters from this experiment ($D$=-2.1 meV, $J$=1.15 meV, $E$=0.31 meV and $B$=2 T) in our model 
there would be additional steps in dI/dV at bias voltages $V$=1.6 mV and $V$=4 mV with heights on the order of 0.01-0.02 $\mu S$. Looking at the two uppermost curves in Fig. 2(c) in Ref.~\cite{yan15}
a small steplike feature does seem to be present at each of these voltages. These possible hints at in-gap $dI/dV$ features of course 
need to be checked - whether they exist could e.g. be verified by data for larger values of transverse anisotropy strength, which we believe would be very interested to investigate.

Finally, interesting questions for future research are to study the effect of transverse anisotropy on 
non-equilibrium spin dynamics in chains of magnetic atoms, on dynamic spin phenomena such as the formation of e.g. magnons, spinons and domain walls,
and on switching of N\'eel states in antiferromagnetically coupled chains.
We acknowledge valuable discussions with F. Delgado and A.F. Otte.
This work is part of the research programme of the Foundation for Fundamental Research on Matter (FOM), 
which is part of the Netherlands Organisation for Scientific Research (NWO). 
\appendix
\section{Equilibrium populations ${\bf P_M}(V)$}
\begin{figure*}[ht]
\centering
\begin{minipage}{0.19\textwidth}
  \centering
  \includegraphics[height=3.3cm,width=1.\linewidth]{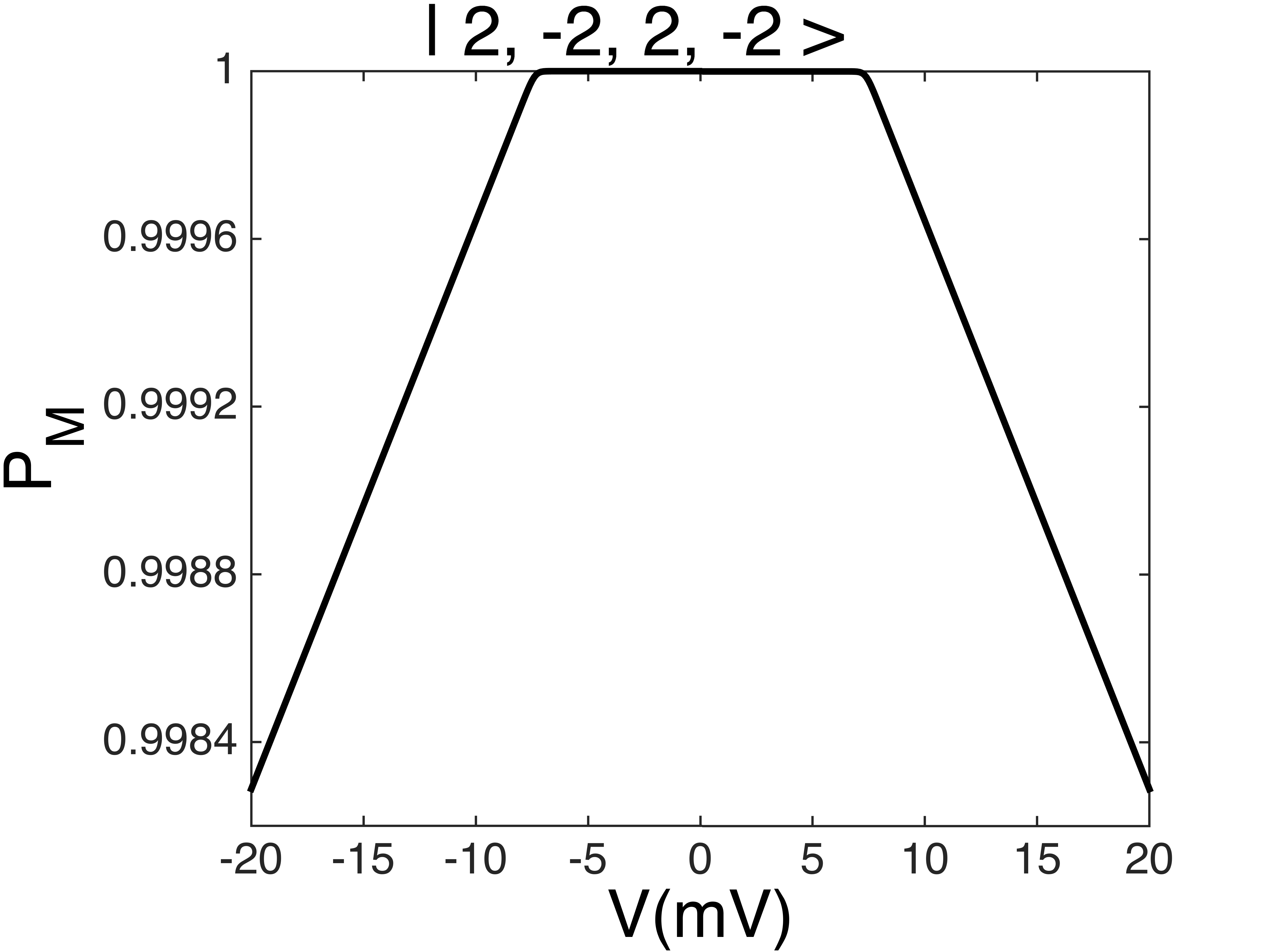}
  \label{fig:test1}
\end{minipage}
\begin{minipage}{.19\textwidth}
  \centering
  \includegraphics[height=3.3cm,width=1.\linewidth]{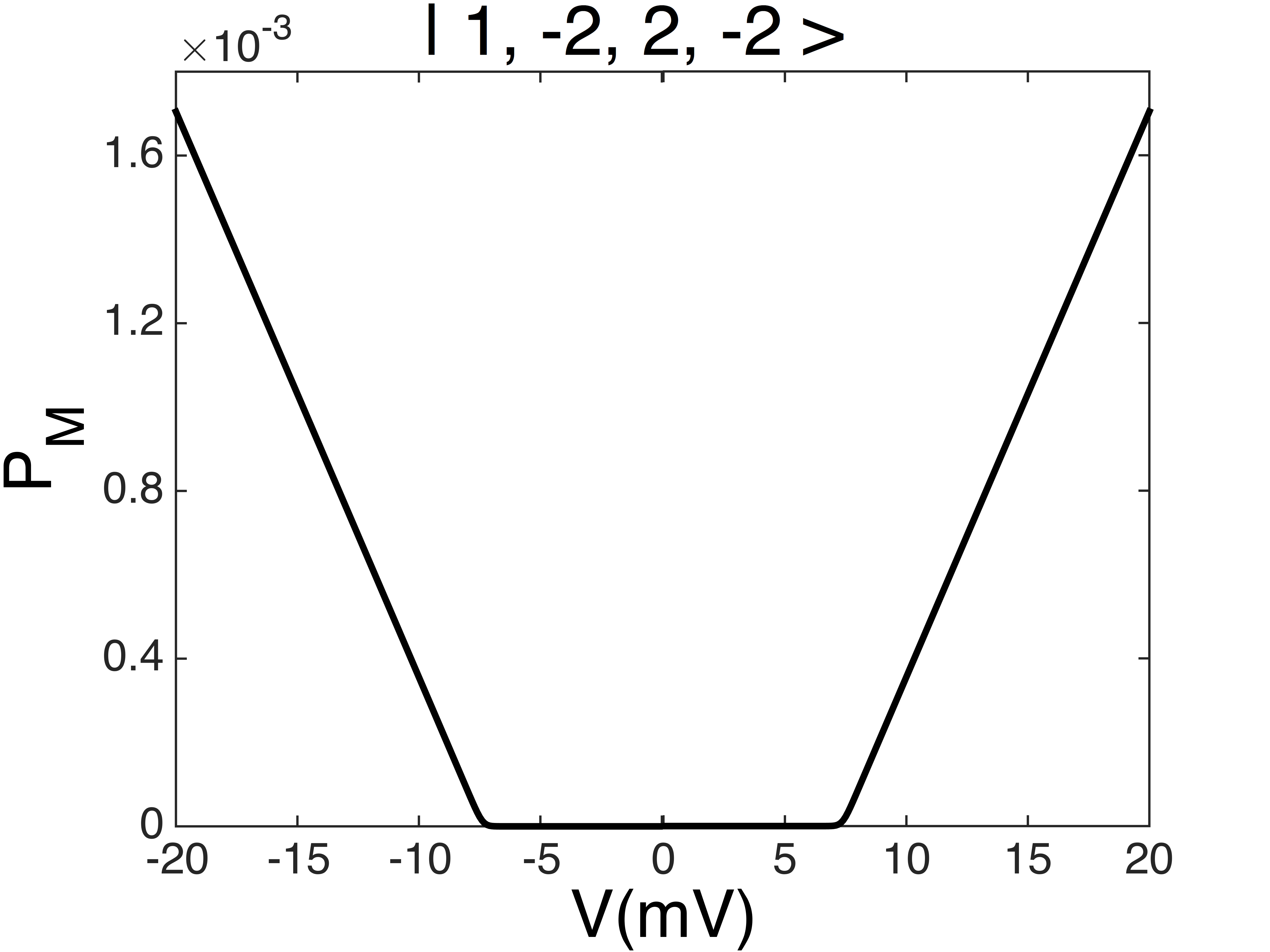}
  \label{fig:test2}
\end{minipage}
\begin{minipage}{.19\textwidth}
  \centering
  \includegraphics[height=3.3cm,width=1.\linewidth]{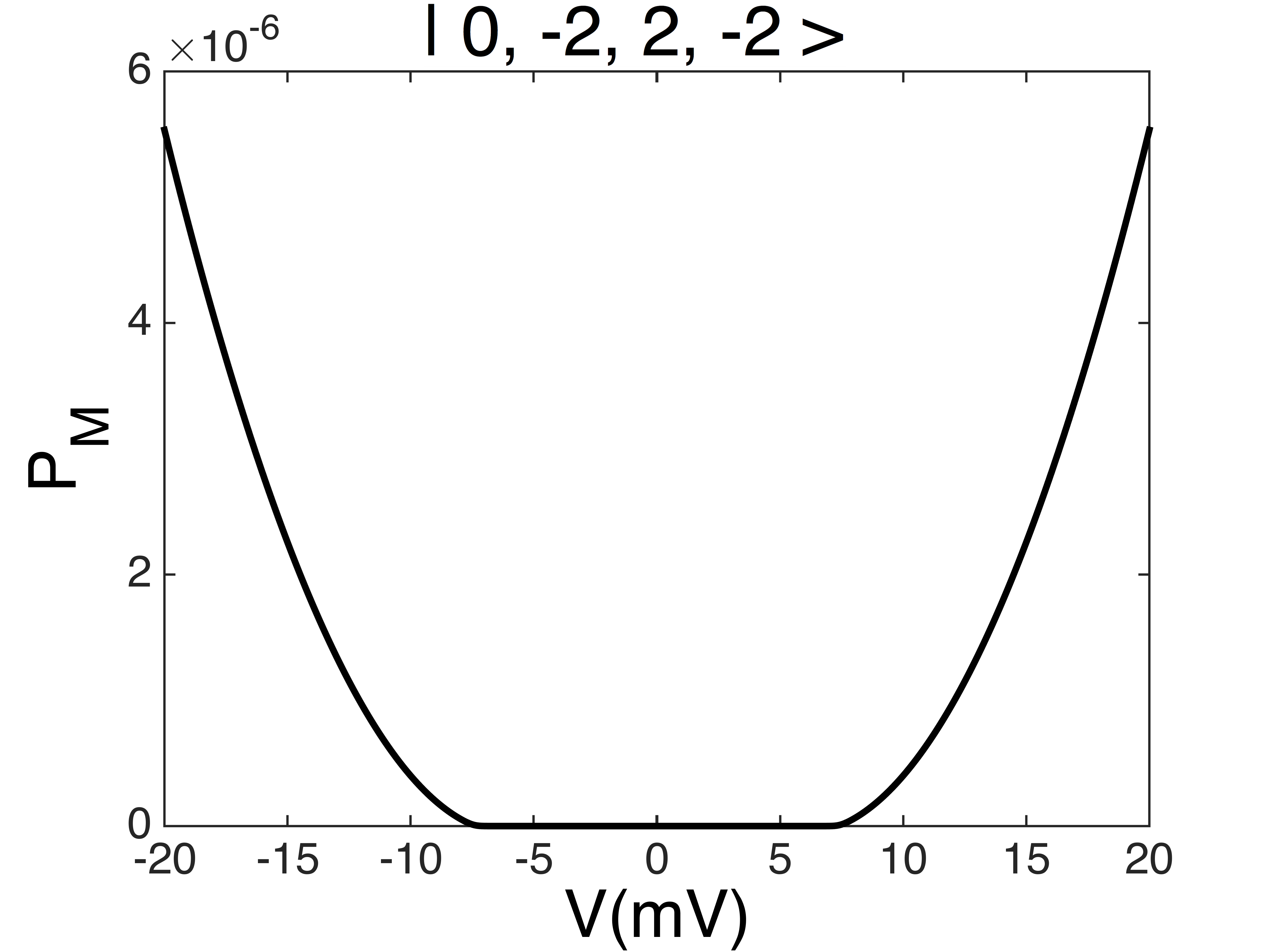}
  \label{fig:test3}
\end{minipage}
\begin{minipage}{.19\textwidth}
  \centering
  \includegraphics[height=3.3cm,width=1.\linewidth]{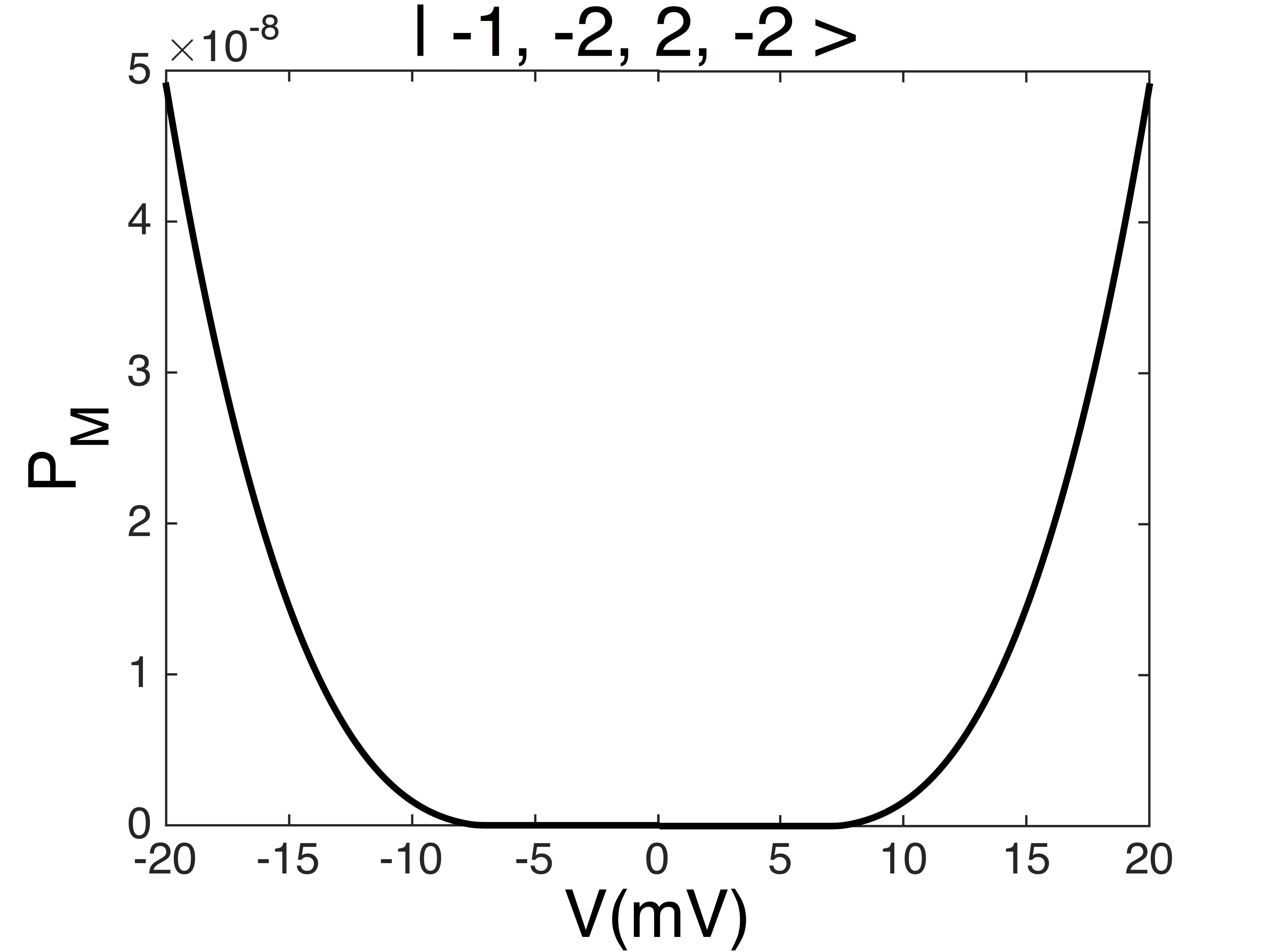}
  \label{fig:test3}
\end{minipage}
\begin{minipage}{.19\textwidth}
  \centering
  \includegraphics[height=3.3cm,width=1.\linewidth]{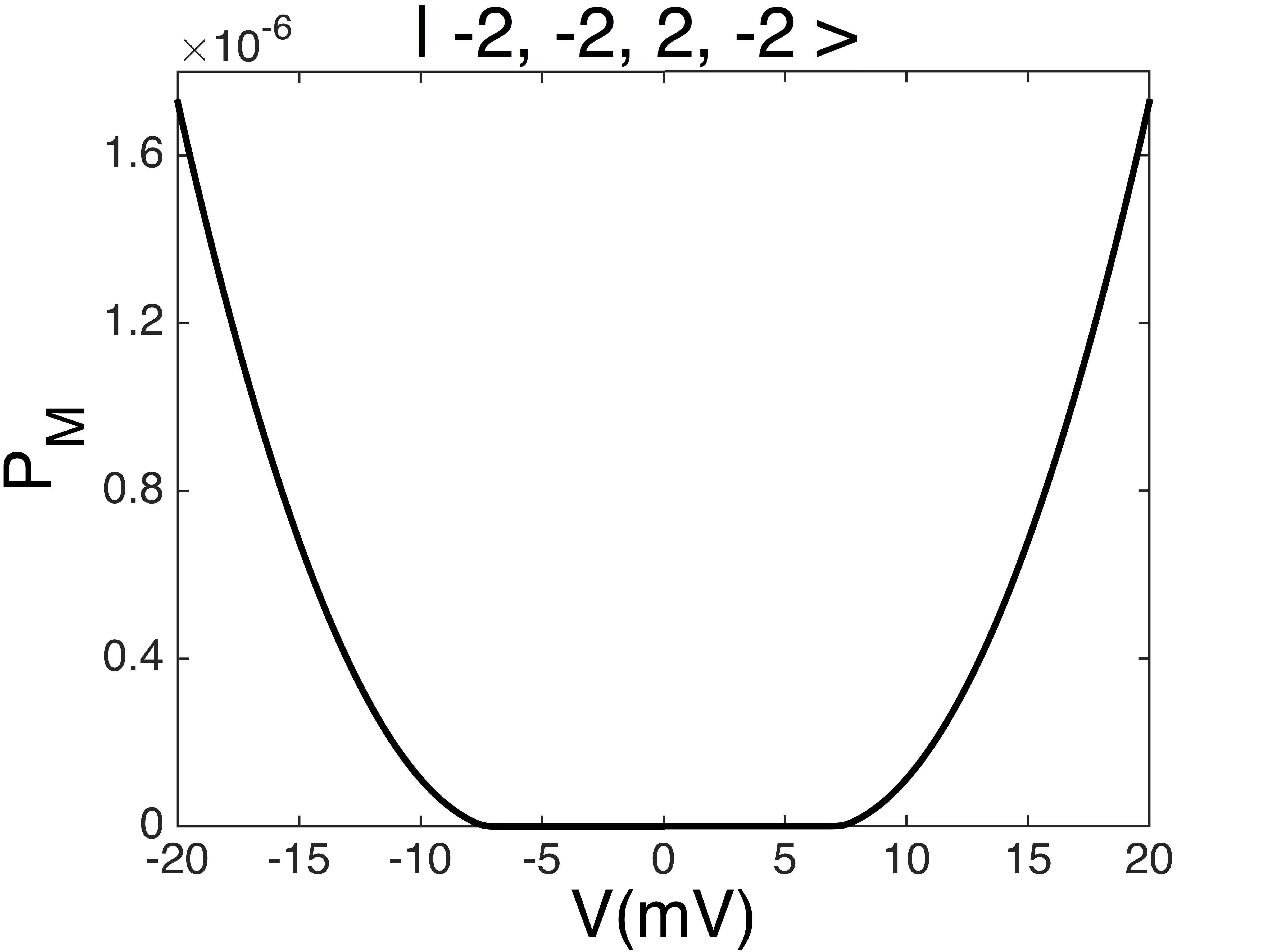}
  \label{fig:test3}
\end{minipage}
\caption{Steady-state population $P_M(V)$ as a function of the applied bias voltage of the five eigenstates with $m_1 \in [2,..,-2]$ of a 
four-atom antiferromagnetic chain in the ground state (see also Fig.~\ref{fig:DIDVAFM}). Parameters used are $D=-1.3$\, meV, $J=1.7$\, meV, $B=1$\, T and $T=1$\, K.}
\label{fig:PM}
\end{figure*}
\noi In this appendix we derive and analyze an analytic expression for the equilibrium population $P_M(V) \equiv P_{m_1,\ldots,m_N}(V)$ for $E=0$
(here the label $M=(m_1,..,m_N)$ refers to the quantum state $|m_1,..m_N\ra^{0)}$).
When calculating and plotting the current $I(V)$ [Eq.~(\ref{eq:current})] 
up to lowest order in $E$, we calculate and include $P_M(V)$ up to lowest order in $E$. 
 As the resulting expressions for $P_M(V)$ are rather lengthy, we do not include them here, but instead show $P_M(V)$ for $E=0$, in order to provide analytical insight
 into the dependence of $P_M$ on the applied bias voltage $V$. We have verified that in the voltage bias range considered in this paper ($\sim 40$~meV) the effect of including nonzero transverse 
 anisotropy in the master equations (\ref{eq:master2}) below is small ($<3\%$), so that the populations $P_M(V)$ are to a reasonable approximation
 given by the solution (\ref{eq:Psol}) for $E=0$.
$P_M(V)$ is the steady-state solution of the master equation~\cite{delg10,delg11}:
\bea
\frac{dP_M(V)}{dt} & = & \sum_{M^{\prime}} P_{M^{\prime}}(V)W_{M^{\prime},M}(V)\ - \nn \\ 
& & \hspace*{0.5cm} P_M(V) \sum_{M^{\prime}} W_{M, M^{\prime}}(V) = 0,
\label{eq:master2}
\eea
with
\bea
W_{M, M^{\prime}}(V) & = & \sum_{\stackrel{\eta=S}{\eta^{\prime}=S,T}} W_{M, M^{\prime}}^{\eta \to \eta^{\prime}}(V) \nn \\
& = & W_{M, M}^{{\rm S}\, \to \, {\rm T}}(V) 
+ W_{M, M^{\prime}}^{{\rm S}\, \to \, {\rm S}} + W_{M, M^{\prime}}^{{\rm S}\, \to\, {\rm T}}(V). \nn \\
\label{eq:rates} 
\eea
The master equation~(\ref{eq:master2}) was derived in Ref.~\cite{delg11} and relies on the assumptions that $\hbar/(k_B T)$, $\hbar/(E_n - E_{n^{\prime}}) \ll 1/W_{n,n^{\prime}}$, 
where $\hbar/(k_B T)$ is the correlation time of the electrons in the leads, $\hbar/(E_n - E_{n^{\prime}})$ the period of coherent evolution and $W_{n,n^{\prime}}$ the scattering time. 
The three rates in Eq.~(\ref{eq:rates}) $W_{M, M}^{{\rm S}\, \to \, {\rm T}}(V)$ (no induced spin flip; spin-independent contribution to the elastic current),
$W_{M, M^{\prime}}^{{\rm S}\, \to \, {\rm S}}$ (spin flip, but no contribution to the current) and $W_{M, M^{\prime}}^{{\rm S}\, \to\, {\rm T}}(V)$ (spin-flip, contribution to the inelastic
current) are given by Eqns.~(\ref{eq:WW1})-(\ref{eq:WW3}) - see also the discussion of these rates and their dependence on $V$ at the beginning of section~\ref{subsec-transitionrates}. 
We assume the tip-induced exchange couplings $W_{m_1,..,m_N^{\prime}}^{{\rm T}\, \to \, {\rm T}}$ and 
$W_{m_1,..,m_N^{\prime}}^{{\rm T}\, \to \, {\rm S}}$ 
to be negligible for $V>0$ as observed in experiments on atomic spin chains~\cite{spin14} - these rates could, however, straightforwardly be included. 
Assuming the STM tip to be only coupled to the atom at site $j$, taking $E=0$, 
and writing $P_{m_j}(V) \equiv P_{m_1,..,m_j,..,m_N}(V)$, Eq.~(\ref{eq:master2}) can be written as
\bea
0 & = & H_{m_j}(V) P_{m_j + 1}(V)\ +\  K_{m_j}(V) P_{m_j - 1}(V) \nn \\
& & \ - \ L_{m_j}(V) P_{m_j}(V).
\label{eq:diffeq}
\eea
The solution of Eq.~(\ref{eq:diffeq}) is given by:
\bea
P_2(V) & = & \frac{K_2(V) K_1(V) K_0(V) K_{-1}(V)}{N(V)}\, T_j \nn \\
P_1(V) & = & \frac{H_1(V) K_1(V) K_0(V) K_{-1}(V)}{N(V)}\, T_j \nn \\
P_0(V) & = & \frac{H_1(V) H_0(V) K_0(V) K_{-1}(V)}{N(V)}\, T_j \label{eq:Psol} \\
P_{-1}(V)& = & \frac{H_1(V) H_0(V) H_{-1}(V) K_{-1}(V)}{N(V)}\, T_j \nn \\
P_{-2}(V) & = & \frac{H_1(V) H_0(V) H_{-1}(V) H_{-2}(V)}{N(V)}\, T_j, \nn
\eea
with
\bea
N(V) & \equiv & H_1 H_0 H_{-1} H_{-2} + H_1 H_0 H_{-1} K_{-1} + \nn \\ 
& & H_1 H_0 K_0 K_{-1}\, +\, H_1 K_1 K_0 K_{-1} + \nn \\
& & K_2 K_1 K_0 K_{-1} 
\nn \\
H_{m_j}(V) & = & W_2\, F_{m_j,-}(0)\, +\,  W_3\, F_{m_j,-}(V)  \nn \\ 
K_{m_j}(V) & = & W_2\, G_{m_j,-}(0)\, +\, W_3\, G_{m_j,-}(V) \nn \\
L_{m_j}(V) & = & 
W_2\, [F_{m_j,+}(0) + G_{m_j,+}(0)]\, + \nn \\
& & \, W_3\, [F_{m_j,+}(V) + G_{m_j,+}(V)] \nn \\
F_{m_j,s}(V) & \equiv &  C_{m_j}\, \frac{eV + s \Del_{m_j,F}}{1 - \exp{[ - \beta (eV + s \Del_{m_j,F})] }}
\nn \\
G_{m_j,s}(V) & \equiv &  C_{m_j-1}\, \frac{eV - s \Del_{m_j,G} }{1 - \exp{[ - \beta (eV - s \Del_{m_j,G} )]}} 
\nn \\
\Del_{m_j,F} & \equiv & E_{m_1,..., m_j,..,m_N} - E_{m_1,..., m_j+1,..,m_N} \nn \\
& = & -(2 m_j + 1) D\, -\, (m_{j-1} + m_{j+1}) J\, +\,  E_Z 
\nn \\
\Del_{m_j,G} & \equiv & E_{m_1,..., m_j-1,..,m_N}- E_{|m_1,..., m_j,..,m_N} \nn \\
& = & -(2 m_j - 1) D\, -\, (m_{j-1} + m_{j+1}) J\, +\, E_Z 
\nn
\vspace*{0.5cm}
\eea
and $C_{m_j}$ given by Eq.~(\ref{eq:Cmj}). It is straightforward to verify that Eq.~(\ref{eq:Psol}) fulfills conservation of population: 
$\sum_{m_j=-2}^2 P_{m_j} (V) \equiv T_{m_j}(V)$, where $T_{m_j}(V) \equiv T_{m_1,..,m_j,..m_N}(V)$ $\in [0,1]$ denotes the total (time-independent) spin population of site $j$ for a given
set of values $m_i$ at the other sites $i \neq j$, and $\sum_j T_j(V)=1$.
Fig.~\ref{fig:PM} shows the equilibrium population $P_M(V)$ [Eq.~(\ref{eq:Psol})] for a chain of four antiferromagnetically coupled atoms with population initially
in the ground state $|2,-2,2,-2\ra^{(0)}$ and the STM tip coupled to the first atom. We see that for this initial state and $T_{m_1}(V)=1$, corresponding 
to low temperatures $k_B T \ll |\Del_1|$ initially all population is located in the ground state (for larger $k_BT > |\Del_1|$ more levels can become occupied, 
when different $T_j$ start to contribute). The ground state population decreases starting at $eV= \pm |\Del_1|$
when the chain can make a transition to the first excited state. However, the population decrease is only slight: at all voltages considered the equilibrium population of the first excited state 
(or higher excited states) is at least a factor $10^3$ 
smaller than the probability of the chain being in the ground state.

% References


\begin{thebibliography}{99}
\bibitem{jaco15}
See also, in a different context, D. Jacob and J. Fern\'andez-Rossier, ArXiv:1507.08474, in which the effect of anisotropy on the
Kondo effect is discussed.
\bibitem{step}
The steplike increase at $eV=|\Del_2|$ is not included in the analytic expression (\ref{eq:currentNeel}), where we assumed $P_{|-2,2,..,-2,2\ra^{(0)}}=1$. Substituting instead the exact 
stationary solution of $P_M$, Eq.~(\ref{eq:Psol}), into
Eq.~(\ref{eq:current}) is straightforward and does lead to this step. However, since the resulting expression is somewhat lengthy we have omitted it here.
\end{thebibliography}
\end{document}